# Radiation Chemistry of Organic Liquids: Saturated Hydrocarbons.

Ilya A. Shkrob, Myran C. Sauer, Jr., and Alexander D. Trifunac.

*Chemistry Division , Argonne National Laboratory, Argonne, IL 60439*

**Abstract.**

In this review, several problems of pivotal importance to radiolysis of saturated hydrocarbons are examined. Special attention is paid to the chemistry of radical cations, high-mobility holes, excited state and spur dynamics, and magnetic field and spin effects and optically detected magnetic resonance spectroscopy.

## 1. INTRODUCTION

In this review, we examine radiolysis of neat organic liquids. The better studied and most common organic solvents are saturated hydrocarbons and alcohols. By virtue of having low dielectric constant and only C-C and C-H bonds, hydrocarbons represent an ideal medium to examine the fundamental mechanisms of radiolysis in non-polar media. For lack of space, the discussion will be limited to paraffins, branched alkanes, and cycloalkanes.

Many studies of saturated hydrocarbons have been carried out over the years; we refer the reader to previous reviews by Hummel [1], Swallow [2], and Holroyd [3]. Here we focus mainly on the last decade and discuss some of the more recent studies. Not too surprisingly, many issues examined in this chapter relate to the topics addressed in our own work; we apologize for this deficiency and hope that other reviews in this volume would complement this chapter. Specifically, we concentrate on the early stages of radiolysis and exclude from our scope chemical transformations of secondary radiolytic products, in particular, those derived from the solutes. We also limit our examination to low-LET radiation, such as UV and VUV photons, x- and γ- rays, and fast electrons. Significant progress has been made in understanding the chemistry in the high-LET tracks and the omission is only due to the lack of space. Finally, we do not review the studies on dynamics and chemistry of electrons. Again, this is not to ignore excellent studies carried out over the last decade. However, our emphasis is on the bond-breaking chemical reactions. In radiolysis of hydrocarbons, these mainly occur in the excited states and solvent holes.

A brief sketch of radiation events in hydrocarbons is appropriate at this point. The ionizing radiation interacts with the solvent to produce excited solvent molecules and electron-hole pairs. Spurs containing one or several electron-hole pairs are generated in scattering events involving the primary and secondary electrons. A large fraction of the ejected electrons thermalize before they escape the Coulomb attraction of the positive charges. Few of these electrons can escape beyond the Onsager radius (at which the Coulomb attraction between the charges is ~ kT) and the majority recombine with the holes. The distance an electron travels in the medium reflects its kinetic energy and the thermalization properties of the medium. Being non-polar, hydrocarbons are unable to "solvate" the electrons as happens in polar liquids (water, alcohols) and localized electrons remain close to the conduction band. Thermal promotion of these electrons to the conduction band leads to overall high mobility. As a result, rapid charge recombination is a dominant feature of the radiation chemistry of hydrocarbons and represents a "clock" against which all other processes compete.

The recombination of electron-hole pairs yields energetic, unstable solvent excited molecules. These excited states and pre-thermalized charges are surrounded by the solvent, and the fate of the excitation is determined by the deactivation, fragmentation, and chemical reactions of these short-lived species with the surrounding molecules. Understanding these rapid processes is the most important problem in the radiation chemistry. Unfortunately, the last decade witnessed little progress in this field. This impasse may continue since even the nature and chemistry of the lower excited states and relaxed solvent holes remains poorly understood. We believe that solving several long-standing problems in the chemistry of these species was the main achievement in the last decade. These studies provide a beachhead for the attack on the final frontier: the chemistry of highly excited neutral and charged states. Judging from the past experience, solving this "radiation chemistry" problem will require going far beyond radiation chemistry methods such as pulse radiolysis. This is why the ultrafast laser studies on ionization and excitation in hydrocarbons, started in this decade, are so important.

While one may dream that developments in other fields would bring closure to some of the "radiation chemistry" problems, other problems can be solved only by radiation chemists themselves. Among the latter problems is the modeling of complex spur reactions. In multiple-pair spurs, many reactions occur simultaneously, resulting in tangled, inhomogeneous kinetics: charges recombine, excited solvent molecules fragment yielding radicals and neutrals, these

2.

species react with the primary ions and each other, the secondary ions react with the products, etc. Due to the high density of the reactive species, such transformations may occur in mere nanoseconds. This complexity makes the analysis of the early events a challenging task. Any understanding of the radiation chemistry will require adequately complex models of the spur processes. Development of such models has progressed significantly in this decade. Though the current computer models are still crude and primitive, they provide a benchmark against which theoretical ideas may be tested. These modeling efforts have already revitalized old discussions about the nature of radiolytic spurs. Without doubt, the future belongs to the more involved and rigorous models and, eventually, to the first-principle calculations. At the present, however, even the most elaborate computer models are as good as the parameters fed into them, and many of these parameters have not been experimentally measured.

This chapter is organized in the following way. First, studies on high-mobility solvent holes in cyclic alkanes are discussed. This topic unifies many issues in the chemistry of the solvent holes; these are examined along the way. Second, we summarize the recent advances in understanding the spur chemistry, starting from the laser and VUV studies of isolated ion pairs to modeling the multiple-pair spurs. The origins of spin effects and the importance of these effects for radiolysis are discussed in the same section. Third, we examine the studies on solvent and solute excited states in radiolysis of hydrocarbons. In the concluding section, we outline the most important problems to be addressed in the forthcoming decade.

## 2. SOLVENT HOLES WITH ANOMALOUSLY HIGH MOBILITY

*2.1 High mobility cations.* Ionization of several cycloalkane liquids - cyclohexane, methylcyclohexane, *trans*-decalin and *cis*-decalin - produces cations whose drift mobilities are 5-to-25 times greater than the mobilities of normally-diffusing molecular ions and (in some cases) thermalized electrons in these liquids [4-8]. These high-mobility cations are shown to be cycloalkane solvent holes with unusually long natural lifetimes (0.2 μs to 5 μs). This long lifetime and the high mobility of the cycloalkane holes makes it possible to study their reactions using microwave [7-9] or direct current (dc) [10-19] conductivity, an option that does not exist for other hydrocarbons. Indeed, in room-temperature paraffins, the solvent holes have natural lifetimes ranging from 1 ns ($C_5$) to 33 ns ($C_{16}$) [20,21] due to rapid dissociation of the C-C or C-H bond(s) and deprotonation,

$$RH^{\bullet +} + RH \longrightarrow R^{\bullet} + RH_2^+ \qquad (1)$$

Similarly short lifetimes are expected for branched alkanes, such as isooctane [22]. Due to these lifetime limitations, the chemical behavior of cycloalkane holes is understood in more detail than that of the solvent holes in other hydrocarbon liquids.

From conductivity studies, it is known that the cycloalkane holes rapidly react with various solutes, typically by electron or proton transfer [7-19]. These scavenging reactions establish the identity of the high-mobility cations as the solvent holes: Rapid generation of aromatic radical cations ($A^{•+}$) in reactions of the holes with aromatic solutes (A) was observed using pulse radiolysis - transient absorption spectroscopy [4,5,6,20,23-25] and, more recently, using pulse-probe laser-induced dc conductivity [26]. Rapid decay of the conductivity and transient absorbance signals from the cycloalkane holes was also observed [4-25].

It has long been speculated that the high-mobility solvent holes exist in hydrocarbons other than the four cycloalkanes. Recently, high-mobility solvent holes were observed in 2,6,10,15,19,23-hexamethyltetracosane (squalane) [24] and in cyclooctane [27]. In the squalane, rapid electron-transfer reactions of solvent holes with low-IP solutes were observed using transient absorbance spectroscopy and magnetic resonance [24]. Fast diffusion and high-rate scavenging reactions of the squalane holes were also observed using magnetic level-crossing [28,29] and quantum beat [29] spectroscopies (see reviews [30] and [31] for the principles of these recently developed spin coherence spectroscopies). Rapid scavenging reactions were also found to account for the anomalies in the magnetic field effect observed for delayed solute fluorescence induced by VUV excitation of squalane [32,33]. In cyclooctane, high-mobility solvent holes were observed using time-dependent electric-field-modulated delayed fluorescence [27] (with this technique, the electric field is used to sort the radical ions by their drift mobilities [34]). This study supports previous observation of rapid scavenging of cyclooctane holes by aromatic solutes in the initial stage of radiolysis [35]. Both in the squalane and cyclooctane, the natural lifetime of the high mobility solvent hole is less than 20 ns [24,27-35]. It is still unclear whether this lifetime is limited by the inherent instability of the solvent holes or by their reactions with impurity. Perhaps, future work will reveal more examples of such short-lived high mobility holes. There is a recent suggestion of the presence of such holes in cyclopentane and cycloheptane; their natural lifetimes must be less than 5-10 ns [27]. Faster-than-normal scavenging of short-lived isooctane holes by diphenylsulfide and biphenyl was observed using quantum beat [36] and transient absorption spectroscopy [20]. Therefore, in addition to a few cycloalkane liquids that yield *long-lived* high-mobility solvent holes there may be many more hydrocarbon liquids that exhibit *short-lived* high-mobility holes. Importantly, not all hydrocarbon liquids can yield the high mobility solvent holes: Neither short-lived nor long-lived high-mobility cations have been found in linear alkanes.

*2.2 The solvent holes.* Previous suggestions that high-mobility cations are proton adducts or carbonium ions have been abandoned [37]. These suggestions originated by consideration of anomalous chemical and physical properties of cyclohexane and *trans*-decalin holes [38]. One of the peculiarities is that these solvent holes, while being paramagnetic species, were not observed by magnetic resonance techniques, both in neat cycloalkanes and in dilute solutions of cycloalkanes in high-IP liquids [39-48]. Instead, the resonance signals from olefin radical cations were observed (see below) [42,44,48]. This suggested a short life time for the solvent holes. The kinetic data on delayed fluorescence [49] and transient absorbance [6,20,25] in radiolysis of cyclohexane supported this conclusion. It was concluded that the lifetime of the solvent hole is only 20-30 ns [6,25], or 10 times shorter than the lifetime of the high-mobility cations observed in the conductivity experiments. Only recently was it determined what causes the appearance of the lifetime-limited kinetics in the pulse radiolysis experiments: Some impurities in cyclohexane reversibly trap the solvent holes reducing their lifetime [11-14,25].

Another crucial finding was the realization that rapid spin-lattice $(T_1)$ relaxation in the high-symmetry cycloalkane radical cations precludes their detection with optically-detected magnetic resonance (ODMR) [39-48], the technique which was routinely used to study radical cations in radiolysis of hydrocarbons [38, 50]. For example, *trans*-decalin•+ isolated in room-temperature cyclohexane has $T_1 \approx 7$ ns [50] while typical solute radical cations have $T_1 \sim 1$ µs. Since it takes several tens of nanoseconds to flip the electron spin with the microwave radiation (which is required for the magnetic resonance detection) radical cations of some cycloalkanes cannot be detected by ODMR.

The ultrafast spin-lattice relaxation is caused by dynamic averaging between the ground and excited states of the radical cations. The near degeneracy of the lower two states results through the Jahn-Teller distortion of highly symmetric radical cations [51,52]. The gap between the two lower states is greater for methylcyclohexane•+ and *cis*-decalin•+ and these two radical cations exhibit more regular magnetic resonance behavior. For cyclohexane•+ in hydrocarbon matrices, the dynamic averaging is so efficient that using ODMR this radical cation cannot be observed even at 4 K [39-41, 45-47, 50]. In *trans*-decalin•+, the quasi-degenerate states are very close in energy (ca. 0.43 eV in the gas phase [52]) and van der Waals interaction with the host can stabilize the "excited" $^2B_g$ state instead of the "ground" $^2A_g$ state (see discussion in references [50,52]). Therefore, for these high-symmetry radical cations no distinction exists between the "ground" and "excited" states, especially at 300 K. For less symmetric radical cations, the state mixing was not observed.

Radiolytically-generated solvent holes have initial excess energy of several electron-volts. It is generally believed that these excited species relax to the "ground" state on a picosecond time

5.

scale or even faster [37,38,53]. Nevertheless, some authors suggest that certain excited cycloalkane holes have lifetimes in nanoseconds [54,55]. Such suggestions are not completely groundless: in conformationally-hindered species, the structural relaxation may take a long time. For example, the twist-boat to chair transition in room-temperature cyclohexane occurs on a microsecond time scale. Perhaps, the best supported claim for such a long-lived excited solvent hole is found in the study of geminate pair dynamics in methylcyclohexane at 143 K [55,56]. Using transient absorption spectroscopy, it was concluded that the (high-mobility) solvent hole [56] has a (high-mobility) precursor with the natural lifetime of 300 ns [55]. The decay of the precursor can be accelerated upon addition of N$_2$O; without the quencher, the precursor either fragments yielding methycyclohexene radical cations (90 %) or relaxes (10%) [55].

It is difficult to assess the plausibility of this scenario because the data allow for more than one interpretation. N$_2$O rapidly scavenges thermalized electrons and quenches the solvent excited states thus reducing the yield of olefins (that form by the fragmentation of these excited states) [1]. Since in some hydrocarbons the olefin radical cations may be formed in reactions of the solvent holes with the olefins in spurs (see below), the yield of these cations will decrease in the presence of N$_2$O. Therefore, the changes observed upon the addition of N$_2$O are not a clear-cut evidence for the involvement of the excited solvent holes.

The optical absorption spectra of the high mobility solvent holes resemble those for the radical cations isolated in freon matrices [20,22-25]. All of these spectra are bell-shaped featureless curves with maxima in the visible and/or near IR regions. In pulse radiolysis studies, the absorption signal from the solvent hole always overlaps with the signals from the fragment (and/or secondary) radical cations ("satellite ions"), even at the earliest observation times [22-25,57]. Therefore, complex deconvolutions are needed to extract the spectra of the solvent holes. This leaves large uncertainty as for the exact shape of the absorption spectra and the extinction coefficients.

*2.3 The mechanism for the high mobility.* In the early studies, the high-mobility cycloalkane holes were viewed as radical cations that undergo rapid resonant charge transfer [8]:

RH•$^+$ + RH 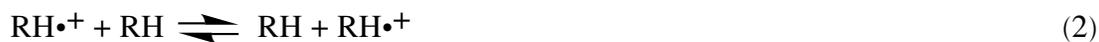 RH + RH•$^+$           (2)

At any given time, the positive charge was assumed to reside on a single solvent molecule and, once in 0.5-2 ps, hop to a neighboring molecule. Reaction (2) was believed to have low activation energy due to similarity between the shapes of cycloalkane molecules and their radical cations.

6.

This model is consistent with many observations. Dilution of cycloalkanes with high-IP alkanes (or higher-IP cycloalkanes) results in the decrease in the hole mobility that correlates with the mole fraction of the cycloalkane in the mixture: the hopping rate decreases when the density of the like molecules decreases [8,14]. The activation energies for the hole mobility, as estimated from conductivity data range from -(3±1) kJ/mol for *trans*-decalin and cyclohexane to (6-7) kJ/mol for methylcyclohexane and *cis*-decalin [7,8,10]. The activation energies for the highest-rate scavenging reactions range from 4 kJ/mol to 9 kJ/mol [10]. Apparently, the migration of the solvent hole requires little thermal activation.

The occurrence of reaction (2) is firmly established experimentally. Charge transfer between *c*-$C_6D_{12}^{\bullet+}$ and *c*-$C_6H_{12}$ was observed in the gas phase, where it proceeds at ~ 1/3 of the collision rate [58]. Reaction (2) was observed for radical cations and molecules of *cis*- and *trans*-decalins in dilute cyclohexane solutions (where it proceeds with a diffusion-controlled rate) [50,59]. In low-temperature solid hydrocarbons (4-30 K), the hole hopping due to reaction (2) may be observed through the time evolution of the resonance lines in ODMR spectra [39,45]; the residence time of the self-trapped holes at a given molecule is 0.1-1 μs. At higher temperatures, the spectral diffusion caused by the rapid reaction (2) causes the ODMR spectrum to collapse to a single narrow line. For solvent holes in liquid *cis*-decalin, *trans*-decalin, and squalane, this narrow line was observed, albeit indirectly, using magnetic level-crossing and quantum beat spectroscopies [28,29,50,59]. In squalane, the residence time of the hole at the solvent molecule is relatively long, ca. 0.2-0.3 ns [24,28,29], and any mechanism of the hole migration other than reaction (2) is implausible. Thus, there is little doubt that reaction (2) occurs in liquid hydrocarbons. Nevertheless, it is doubtful that this reaction *per se* is the cause for the high mobility of the solvent holes in cycloalkanes.

Indeed, both matrix-isolation EPR experiments and quantum-mechanical calculations indicate that the neutral cycloalkanes and their ground-state radical cations have rather different geometries [50,51,52]. For example, in *cis*- and *trans*-decalins the bridging $C_9$-$C_{10}$ bond elongates from 0.153-0.156 nm in the neutral molecules to 0.19-0.21 nm in the $^2A_1$ and $^2A_g$ states of the radical cations, respectively [10,52]. Upon charging, the molecule undergoes considerable structural and energetical relaxation, losing at least 0.5-0.7 eV. If the charge transfer reaction (2) were a single-step process, it would require the activation energy of 1-2 eV [10].

What makes this resonant charge transfer possible? It was suggested that in the gas phase, reaction (2) proceeds through the formation of a collision complex in which the charge is shared by both of the cycloalkane molecules [58]. This sharing considerably reduces the barriers for the structural relaxation. According to MNDO calculations, the $C_2$-symmetric cyclohexane$_2^{\bullet+}$

7.

dimer in which the charge is shared equally between the cyclohexane monomers, is only 150 meV higher in energy than the state in which the charge is localized on a single monomer [10]. Continuing this line of reasoning, it may be assumed that in liquid cycloalkanes the charge is shared between several solvent molecules and this sharing further reduces the hopping barrier. In such a case, the solvent hole is a small polaron whose rapid migration is caused by phonon-assisted hopping [10].

The formation of the polaron causes delocalization of the hole. Unusually large reaction radii in electron-transfer reactions of the solvent holes were first considered as evidence for such a delocalization [8]. However, more recent measurements of the hole mobility suggest that these radii were overestimated [10,14]. The best evidence for the delocalization of the solvent hole was provided by studies on dc photoconductivity in cyclohexane-methylcyclohexane mixtures [10,11]. While the addition of less than 5-10 vol % of methylcyclohexane reduces both the dc conductivity signal and its decay rate, further addition of methylcyclohexane yields little change in the conductivity signal and kinetics. The initial reduction is accounted for by rapid reversible trapping of cyclohexane holes by methylcyclohexane ($\Delta G^0 \approx$ -0.11 eV [11]). Since the isolated radical cations of methyl-cyclohexane in cyclohexane have normal mobility, the excess conductivity signal is proportional to the equilibrium concentration of the cyclohexane holes. At higher concentration of methylcyclohexane, the equilibrium fraction of the cyclohexane holes becomes lower. When this fraction becomes less than the ratio $\mu_h/\mu_i$ of the mobilities of the high mobility $(\mu_h)$ and normally diffusing $(\mu_i)$ ions, the conductivity should decrease severalfold. This decrease was not observed. We conclude that the migration of methylcyclohexane•+ in 5 vol % methylcyclohexane solution is as rapid as that of the solvent holes in neat methylcyclohexane. When the methylcyclohexane is diluted by *n*-hexane instead of the cyclohexane, the conductivity signal decreases proportionally to the fraction of *n*-hexane. Thus, it was cyclohexane which made the difference.

This result suggests that the methylcyclohexane holes are coupled to the solvent, forming a polaron. This coupling makes the charge migration of methylcyclohexane holes in cyclohexane as efficient as in neat methylcyclohexane. From the critical concentration of methylcyclohexane, the delocalization radius was estimated as ca. 1 nm, or 4 to 5 molecular diameters [10].

*2.4. The formation of high-mobility holes and "satellite ions".* As was briefly mentioned above, radiolysis of hydrocarbons results in the formation of several types of cationic species besides the solvent holes. Most of these "satellite ions" are generated within the first nanosecond after the radiolytic pulse.

8.

Transient absorption spectra of some "satellite ions" closely resemble the spectra of olefin radical cations. In cyclohexane, a band centered at 270 nm (at 2 ns [22]) is observed from 250 ps [25] after the ionization event (this band overlaps with the strong 240 nm band of cyclohexyl radicals [22]). The scavenging behavior and the decay kinetics of the UV-absorbing species suggest that they are normally-diffusing radical cations [25]. In the first few nanoseconds after the ionization event, the VIS absorbance is dominated by solvent excited states [22,57]. When the thermalized electrons are rapidly scavenged using a suitable electron acceptor (halocarbons or $N_2O$), this absorbance is much reduced and, in addition to cyclohexene$^{\bullet+}$, one observes the absorption bands of cyclohexane holes (the latter may be rapidly scavenged using alcohols) and some other cations that absorb in the red [25]. The latter signals are clearly distinguishable as early as 1-5 ns after the radiolytic pulse (at earlier delay times, these signals are swamped by absorption of the excited states). The VIS spectra of the red-absorbing cations resemble those from cyclohexene dimer radical cations or cyclohexadiene radical cations. Thus, shortly after ionization of cyclohexane with fast electrons, at least three types of cation species were observed spectroscopically.

In some cases, the identity of paramagnetic "satellite ions" was established by ODMR [42,44,48]. For example, 9,10-octalin$^{\bullet+}$ was identified in decalins and their solutions [42]. ODMR spectra of "satellite ions" in cyclohexane were related to EPR spectra of matrix-isolated cyclohexene$^{\bullet+}$ (Note that in the liquid cyclohexane, cyclohexene$^{\bullet+}$ undergoes a fast ring-puckering motion that averages hyperfine coupling constants for equatorial and axial protons, so the the EPR spectra of cyclohexene$^{\bullet+}$ in liquid and solid matrices are different) [42,44,48]. In both of these cases, the olefin radical cations were formed in spurs rather than in a reaction of the solvent hole with the olefin in the solvent bulk [42] (octalins gradually accumulate as radiolytic products). Olefin "satellite ions" were also observed in squalane [24].

How do these "satellite ions" form in the early stages of radiolysis ? Two ideas were put forward [1,37,42]. First, the "satellite ions" could be generated by fragmentation of short-lived electronically- and/or vibronically-excited solvent holes formed upon the ionization of the solvent, for example

$$RH^{\bullet+*} \longrightarrow \text{olefin}^{\bullet+} + H_2 \qquad (3)$$

Reaction (3) is exothermic even for the ground-state alkane radical cations (0.4-0.6 eV) but requires overcoming a high potential barrier. Several authors (e.g., [22, 37]) have considered the possibility of deprotonation of the excited hole,

$$RH^{\bullet+*} + RH \longrightarrow R^{\bullet} + RH_2^+ \text{ or } R^{\bullet} + R^+ + H_2 \qquad (4)$$

9.

In cyclohexane and decalins, reaction (1) is endothermic by 0.1-0.4 eV [60] and it seems reasonable that the excitation of the hole may facilitate the proton transfer. Fragmentation of matrix-isolated hydrocarbon radical cations upon excitation with 2-4 eV photons was observed by EPR (see review [61]). For cycloalkanes, the main photoreaction is reaction (3). For radical cations of methyl-branched alkanes, the loss of $CH_4$ was also observed, while the radical cations of linear alkanes prefer to fragment to 2-butene$^{\bullet+}$ and the residual paraffin [60]. Thus, even the lower excited states of some radical cations are dissociative. One suggestion is that the recombination of electronically-excited solvent holes with electrons yields strongly dissociative higher excited states of the solvent [53]. These states are likely to fragment before they relax to the lowest $S_1$ state. In this way, it is possible to explain the low yield of the fluorescent $S_1$ states in some hydrocarbons.

Second, the "satellite ions" could be generated in scavenging reactions of the solvent holes with radiolytic products in multiple-pair spurs [25,61-65]. The olefins are formed upon the fragmentation of excited solvent molecules generated in recombination of short-lived electron-hole pairs [1]

$$RH^{\bullet+} + e^- \longrightarrow {}^{1,3}RH^* \tag{5}$$

$$^1RH^* \longrightarrow olefin + H_2 \tag{6}$$

$$RH^{\bullet+} + olefin \longrightarrow RH + olefin^{\bullet+} \tag{7}$$

This mechanism would also account for rapid generation of carbonium ions in reactions of the solvent holes with radicals [65]

$$RH^{\bullet+} + R^{\bullet} \longrightarrow RH + R^+ \tag{8}$$

The results on low-temperature ODMR suggest that in some hydrocarbons, no "satellite" *radical* cations are formed via the dissociation of excited-state holes. No prompt generation of olefin radical cations was observed in solid paraffins, decalins, and methylcyclohexane at 4-50 K [39-41,45-47]. At these low temperatures, the holes have negligible mobility [39,45] which makes electron transfer reaction (7) slow and inefficient, as it may occur by long-range electron tunneling only. Since the excited-state holes should also be formed in the solid, these ODMR observations suggest that the "satellite" *radical* cations are generated mainly in ion-molecule spur reactions (at least, in some hydrocarbons).

The prompt formation of the "satellite ions" introduces ambiguity in the measurement of the hole mobility in pulse radiolysis. Indeed, the conductivity is a product of the mobility and the yield [8]. The latter quantity is poorly defined since the branching ratio $f_h$ between the high-mobility solvent holes and the "satellite ions" is unknown.

The way around this problem is to generate solvent holes by means other than radiolysis. Since the formation of "satellite ions" is partly due to spur chemistry, in the laser multiphoton ionization of neat hydrocarbons [14-18]

$$RH + n\ h\nu \longrightarrow RH^{\bullet +} + e^- \qquad (9)$$

(that yields isolated electron-hole pairs only) the yield of "satellite ions" would be much reduced. Another way of generating the solvent holes is by a valence-band electron transfer to the photoexcited aromatic radical cation ("hole injection") [10-13,26]

$$A^{\bullet +} + h\nu \longrightarrow A^{\bullet +*} \qquad (10)$$

$$A^{\bullet +*} + RH \longrightarrow A + RH^{\bullet +} \qquad (11)$$

The photoexcitation of $A^{\bullet +}$ may be carried out using the same UV pulse that is used to ionize the solute or using a delayed laser pulse of a different color. The latter method is preferable because by using a delay such that only free $A^{\bullet +}$ exist the solvent hole can be generated without a geminate counterion. Therefore, the scavenging kinetics can be disentangled from the geminate recombination. Both of these approaches were used to study long-lived solvent holes in cycloalkanes.

Solvent holes in neat cycloalkanes were generated by multiphoton ionization (3 x 4 eV or 2 x 5 eV) of the solvent at fluxes in excess of 0.01 J/cm$^2$ [15]. In a typical experiment, the laser-induced dc conductivity was measured as a function of the delay time with resolution better than 3 ns. A similar setup was used to observe the dc conductivity in pulse radiolysis with fast 16 MeV electrons [14]. The decay kinetics of solvent holes in cyclohexane and decalins were consistent with the value of $f_h \approx 1$ for multiphoton laser ionization. For cyclohexane, a lower ratio of $f_h \approx 0.5$ was needed to account for the kinetics observed in pulse radiolysis. (Note that these ratios refer to the situation at ca. 10 ns after the ionization event; the conductivity signal of the holes cannot be measured at earlier time). To be consistent with the observations, the simulations required a higher value for the mobility $\mu_h$ of the cyclohexane holes (1.7x10$^{-2}$ cm$^2$/Vs [12,14]) than previously estimated [7,8]. High yield of the "satellite ions" is not a universal property: In radiolysis of decalins, the ratio $f_h > 0.8$-0.9. Apparently, the yield of the "satellite ions" varies apreciably with the hydrocarbon structure.

Knowing the absolute values of $f_h$ is important since using the previous estimates for $\mu_h$ led to unrealistically large reaction radii for reactions of cyclohexane and methylcyclohexane holes with low-IP solutes (2-3 nm!) [8, 14]. These radii suggested extreme delocalization of the solvent hole. Using the correct mobilities reduces these radii to ca. 1 nm which is close to a typical electron-transfer radius in a non-viscous hydrocarbon.

When reaction (11) is induced by the VIS photons (at 2.3 eV), the initial excess energy of the solvent hole is 0.7-1.2 eV (for low-IP solvents such as decalins) and the question may be raised about the occurrence of reaction (4). The pump-probe conductivity experiment outlined in reference [26] shows that after the *trans*-decalin hole produced from triphenylene$\bullet^{+*}$ disappears by reacting with the triphenylene in solution, the ground-state triphenylene$\bullet^{+}$ is completely recovered. The kinetics of the recovery mirrors the decay kinetics of the solvent hole. These observations indicate that no "satellite ions" are formed in reaction (11). Direct photoexcitation of *trans*-decalin holes at 2.3 eV also did not result in the reduction of the conductivity signals from these holes. No cations other than A$\bullet^{+}$ and high mobility solvent holes were found upon the 5 eV photoexcitation of triphenylene in *trans*-decalin, despite the high excess energy in the holes (ca. 3.8 eV) following the UV-photoinduced "hole injection". Apparently, the excited solvent hole in the *trans*-decalin is very stable; even the anticipated deprotonation does not occur. This explains the low yield of "satellite ions" in radiolysis of *cis*- and *trans*-decalins. It is presently unclear what factors control the stability of electronically-excited holes in hydrocarbon solvents.

Quantum yields $\phi_h$ for single-photon "hole injection" for aromatic solutes in *trans*-decalin correlate well with the gas-phase IP of the aromatic solute. For triphenylene$\bullet^{+}$, $\phi_h \approx 0.016$ was obtained for excitation with 5 eV photons (ca. 3 times the value for 2.3 eV excitation) [26].

*2.5. Ion-molecule reactions of high-mobility solvent holes.* There are several classes of such reactions:

(i) fast irreversible electron-transfer reactions with solutes that have low adiabatic IPs (ionization potentials) and vertical IPs (such as polycyclic aromatic molecules);

(ii) slow reversible electron-transfer reactions with solutes that have low adiabatic and high vertical IPs (such as dimethylcyclopentanes);

(iii) fast proton-transfer reactions;

(iv) slow proton-transfer reactions that occur through the formation of metastable complexes [9,26];

(v) very slow reactions with high-IP, low-PA (proton affinity) solutes.

Rate constants for cyclohexane holes may be found in references [7,8,11,13,14,17], for decalin holes - in references [8,9,12,14,26], for methyl-cyclohexane holes - in references [12,122], for squalane holes - in references [24,30]. The data on the temperature dependence of rate constants of scavenging for the four cycloalkane holes are in reference [10]. For these holes, most of the rate constants were measured by determining the decay kinetics of the transient conductivity signals as a function of the solute concentration. The preferable way of studying the scavenging reactions is by detection of the excess dc conductivity following the "hole injection" reactions (10) and (11) [10-13,26]. In cyclohexane, the determination of the rate constants is complicated by the fact that the solvent hole is in equilibrium with an impurity in the solvent [11].

*Class (i) reactions* were observed in all four cycloalkanes that exhibit long-lived high-mobility holes [4-8,10,13,14,17]. These reactions were also observed in squalane [24, 31] and cyclooctane [26]. The reaction rates linearly scale with the hole mobility as a function of temperature (with exception of *trans*-decalin) and the fraction of cycloalkane in the solvent mixture [14]. The highest rate constants were observed for reactions of cyclohexane hole with low-IP aromatic solutes, $(3-4.5) \times 10^{11}$ $M^{-1}$ $s^{-1}$ at 25°C [13,14]. In these irreversible reactions, a solute radical cation is generated.

*Class (ii) reactions* were directly observed for the solvent holes in cyclohexane [11] and methylcyclohexane [122]. For some solutes (SH), the equilibrium

$$RH^{\bullet +} + SH \rightleftharpoons RH + SH^{\bullet +} \qquad (12)$$

is set on the time scale of the conductivity experiment (> 10 ns). In this case, the decay kinetics of the solvent holes are biexponential. Addition of 1,1-dimethylcyclopentane, *trans*-1,2-dimethylcyclopentane, and 2,3-dimethyl-pentane to cyclohexane or *trans*-decalin, bicyclohexyl, and *iso*-propyl-cyclohexane to methylcyclohexane results in such bimodal scavenging kinetics. The former two cyclopentane derivatives are present as impurity in commercial cyclohexane (10-100 ppm). For addition of methycyclohexane to cyclohexane, the equilibrium (12) is reached so rapidly that the decay kinetics are exponential. Similar rapidly-set equilibria exist for high-mobility holes in mixtures of *cis*- and *trans*-decalins [14,38] and the decalin holes and benzene [26].

The rate constants, $k_{12}$, of the forward reaction (12) are an order of magnitude lower than those of the class (i) reactions, though some of the hole-trapping solutes have comparably low adiabatic IPs. The values of $k_{12}$ did not correlate with the observed $\Delta G^0$ of reaction (12). An explanation was proposed that the rate constants are controlled by the height of the activation barrier determined by the difference in the vertical IP of the solute and the adiabatic IP of the solvent [11]. This suggests that electron transfer to the rapidly-migrating solvent hole (as it passes by the scavenger molecule) is much faster than the relaxation time of the solute radical cations.

*Class (iii) reactions* include proton-transfer reactions of solvent holes in cyclohexane, methylcyclohexane, and squalane [4-8,10,13,14,17,26]:

$$RH + SH \dashrightarrow R\bullet + SH_2^+ \tag{13}$$

The corresponding rate constants are 10-30% of the fastest class (i) reactions and exhibit short reaction radii of 0.15-0.4 nm. Unlike the electron-transfer reactions (that may occur through space), the proton transfer requires close proximity of the donor and acceptor. Thus, short reaction radii of class (iii) reactions suggest a low degree of the solvent hole delocalization.

*Class (iv) reactions* include proton-transfer reactions in *trans*-decalin and decalin mixtures [9,12,14,26]. In neat *trans*-decalin, the reaction rates for high-IP solutes correlate poorly with the solute IP and PA [9, 12]. In binary mixtures of *cis*- and *trans*-decalin, the mobility of the solvent hole linearly scales with the fraction of *trans*-decalin (for which $\mu_h$ is 4.5 times higher than in *cis*-decalin) [14]. While for most of the solutes the rate constants also scale linearly with the hole mobility, for aliphatic alcohols the rate constant systematically decreases with the fraction of *trans*-decalin approaching the value of $(5-6) \times 10^9$ M$^{-1}$ s$^{-1}$ in neat *trans*-decalin [12,14]. This constant is only 10% of the rate constant for alcohols in *cis*-decalin.

Recently, it was demonstrated that the scavenging of *trans*-decalin holes by some alcohols proceeds through the formation of a metastable complex [26]

$$RH + SH \rightleftharpoons \{RH \ldots SH\}^{\bullet+}$$
(14)

$$\{RH \ldots SH\}^{\bullet+} \xrightarrow{\phantom{SH}} \begin{array}{l} R\bullet + SH_2 \\ R\bullet + H^+(SH)_2 \end{array} \tag{15}$$

with a natural lifetime between 24 ns (2-propanol) to 90 ns (*tert*-butanol) at 25°C. In neat decalins, the rate of the complexation is ca. 1/2 of the highest electron-transfer rates (the reaction radius is 0.5-0.7 nm); the overall decay rate is limited by slow proton-transfer reactions (15). The rate constant of unimolecular decay of the complex is $(5-10) \times 10^6$ s$^{-1}$. Though for other class (iii) reactions the bimodality was not observed [9,26], the basic mechanism must be the same. Only for secondary and ternary alcohols is the equilibrium (14) reached so slowly that it can be observed at 25°C on the time scale of the conductivity experiment (> 10 ns). For primary alcohols, the scavenging kinetics are pseudo-first order. However, for $C_3$-$C_6$ alcohols the rate constants do not scale lineraly with the solute concentration, betraying the fast equilibria (14). For these alcohols, PA is relatively low and the decay of the complex is mainly bimolecular. Termolecular reactions analogous to reaction (15) were observed in the gas phase, e.g., for $c$-$C_6H_{12}\bullet^+$ and water [60].

A detailed analysis of the thermodynamics and energetics of reactions (14) and (15) is given in references [26]. Forward reaction (14) has near-zero activation energy; reaction (15) is thermally-activated (20-25 kJ/mol). The stability of the complex increases with the carbon number of the alcohol; the standard heat of the complexation decreases in the opposite direction ($\Delta H_{298}^0$ changes from -39 kJ/mol for ethanol to -25 kJ/mol for *tert*-butanol). Complexes of *cis*-decalin$\bullet^+$ are much more stable than complexes of *trans*-decalin$\bullet^+$ since for the former, the standard reaction entropy $\Delta S_{298}^0$ is 35 J mol$^{-1}$ K$^{-1}$ more positive. The decrease in the entropy is small for both decalins ($\Delta S_{298}^0 > -80$ J mol$^{-1}$ K$^{-1}$), approaching zero for higher alcohols [26]. Similarly small changes in the standard entropy were observed for class (ii) reactions of methylcyclohexane$\bullet^+$ [122]. Since the molecular complex formation could only reduce the degrees of freedom, to account for the small change in the entropy there must be an increase in the solvent disorder. This would be consistent with a hole being a small polaron that orders solvent molecules around it. When the positive charge is compensated, the solvent becomes disordered, and the reaction entropy increases.

Complex mechanism of class (ii) and (iii) reactions may account for the puzzling result in the studies on radiation-induced fluorescence in *cis*- and *trans*-decalins containing 3-100 mM of benzene [54], where it was concluded that on the time scale of geminate recombination of primary pairs in *trans*-decalin (< 1 ns), the hole is scavenged by benzene with rate constant of 7.7x10$^{10}$ M$^{-1}$ s$^{-1}$ (vs. (5-5.5)x10$^9$ M$^{-1}$ s$^{-1}$ observed in the transient conductivity experiments [7,8,12,14]). This was taken as evidence for the involvement of short-lived, reactive excited solvent holes.

It is more likely that the higher rate constant corresponded to the rate constant of forward charge transfer reaction (12) or complexation reaction (14). The gas-phase IP for benzene and *trans*-decalin are very close. In *trans*-decalin solution, $\Delta IP \approx 0.25$ eV, and benzene$\bullet^+$ readily transfers positive charge back to the solvent ($\Delta H^0 \approx -0.27$ eV [26]). This was demonstrated through efficient generation of *trans*-decalin holes by biphotonic ionization of benzene [14] and careful analysis of scavenging kinetics [26]. The decay of the solvent hole is due to slow proton transfer reaction (13) and dimerization of benzene$\bullet^+$ with rate constant of 5.7x10$^9$ M$^{-1}$ s$^{-1}$.

15.

Since benzene$_2^{\bullet+}$ has 0.65 eV lower energy than benzene$^{\bullet+}$, backward charge transfer from the dimer cation is inhibited, and the dimerization shifts the equilibrium (12) to the right side [26]. The forward charge transfer (or complexation) proceeds with rate constant of $(1.1\pm0.1)\times10^{11}$ M$^{-1}$ s$^{-1}$, while $k_{-12} \approx 1.3\times10^8$ s$^{-1}$ (at 25°C) [26]. Therefore, while in the fluorescence studies the solvent holes were observed before the equilibrium (12) was reached (ca. 7 ns in 3 mM solution), in the conductivity studies the solvent holes were observed well after this equilibrium was reached. Characteristically, for the low-IP solute toluene (a class (i) solute), the rate constants measured on short [54] and long time scales [9,12,14] are identical.

Although *class (iv) reactions* were observed for several high-IP, low-PA solutes [8,14,17,19,20], the kinetic data were easy to misinterpret due to traces of low-IP impurity in the inert solute (in particular, for C$_{10}$-C$_{16}$ paraffins [20]). The only reliable data were obtained for scavenging the solvent hole in cyclohexane by cyclopropane [8,17] and for scavenging the solvent holes of cyclohexane and decalins by oxygen [14,19]. For the latter reactions, the reaction constants are $(1-3)\times10^8$ M$^{-1}$ s$^{-1}$ [14], more than two orders of magnitude lower than those for class (i) reactions. Our thermochemical analysis suggests that this reaction is initiated by the H atom transfer to O$_2$ [14]. A possible mechanism for scavenging reaction with cyclopropane is the H$_2^-$ transfer [8].

In concluding this section, we observe that though the nature and the migration mechanism for the high-mobility holes are not yet fully understood, a consistent picture of their chemical, dynamic, and magnetic properties begins to emerge.

### 3. SINGLE-PAIR AND MULTIPLE-PAIR SPURS

*3.1. Single-pair spurs.* Single-photon VUV or multiphoton UV ionization of neat liquid hydrocarbons results in the formation of *isolated* geminate ion pairs, reaction (9). The reaction dynamics in such pairs is less involved than the dynamics in multiple-pair spurs formed in radiolysis. The availability of synchrotron radiation in the 10-40 eV region and short-pulse UV lasers led to a rapid increase in the number of studies on "single-pair spurs". In particular, ultrafast pulse-probe laser spectroscopy was used to study ionization, geminate recombination, and generation of the solvent excited states in neat hydrocarbons [66-69]. This field is still in its infancy: Only in 1997 has the first reliable data on the geminate kinetics of electron-hole pairs been obtained [69]. Eventually, these studies will complement the studies using pulse radiolysis and VUV photoexcitation.

In these ultrafast UV laser studies, the pump energies varied between 4 eV and 5 eV, while the probe pulse energies varied between 0.55 eV and 3.1 eV. Upon biphotonic 5 eV or triphotonic 4 eV excitation, hydrocarbons (such as paraffins, isooctane, cylopentane, cyclohexane, and *trans*-decalin) yield both the solvent excited $S_1$ states and the electron-hole pairs [66]. Questions were raised as to the significance of the ionization channel, since more than 75% of the transient absorbance in the VIS range was from the solvent $S_1$ states [67,68]. It is presently realized that there is a strong spectral overlap between the absorbance signals from these excited states and the primary charge carriers. The $S_1$ state dominates the absorbance in the 1.8-3 eV region [67], while below 0.8 eV, the absorbance is from the electrons only [69]. A biphotonic 3.5-4 eV laser excitation yields the $S_1$ states without the concurrent ionization [70]. Thus, both the $S_1$ state dynamics and the geminate dynamics of single electron-hole pairs may be studied separately using the appropriate excitation and detection conditions. This option does not exist in pulse radiolysis studies.

The results of the ultrafast laser studies are very preliminary. The observed geminate kinetics suggest that the electron thermalization distance distribution in isolated electron-hole pairs can better be described as the exponential than the often assumed $r^2$-Gaussian [65]. This conclusion was supported in the studies on magnetic field dependence of hexafluorobenzene fluorescence in recombination of geminate $C_6F_6^{\bullet -}$ - solvent hole pairs generated upon 10 eV photoionization of isooctane. One study found that even the exponential distribution law was too steep to account for the data and that a power law tail is more appropriate [71] (for radiolytic spurs, Gauss-power distributions had long been advocated [72]). The results on the solvent photoionization are in contrast with the data on multiphoton UV laser ionization of aromatic solutes in the same hydrocarbons that demonstrated good applicability of the $r^2$-exponential or $r^2$-Gaussian distributions for simulation of the geminate pair recombination and free ion yield (for example, [16]). On the other hand, no systematic studies on the pump energy dependence of the geminate kinetics in neat hydrocarbons have been carried out, while it is known from other ultrafast studies (in particular, on photoionization of liquid water [73]) that the thermalization distances strongly depend on the pump energy and the ionization mechanism.

Arguably, the greatest import of the forthcoming ultrafast studies would be assessing the role of short-lived pre-thermalized charges and highly excited states of hydrocarbons. The first step in this direction has been made: It was found that in *n*-hexane and isooctane the electron band in the near IR is shifted to lower energies in the first 2 ps after the ionization event [69]. This finding suggests that the thermalization of electrons in hydrocarbons is not as rapid as generally believed. Therefore, a considerable fraction of reactions in spurs involves pre-thermalized charge carriers.

A recent study [74] demonstrated the potential of femtosecond photoemission spectroscopy to study localization (thermalization) dynamics of electrons in hydrocarbons. The electrons were injected into *n*-heptane bilayer on Ag(111) surface (at 120 K) using a powerful 4 eV pump pulse. The electrons were then ejected to vacuum using a weaker 2 eV probe pulse. The extended and localized electron states were distinguished by the angular dependence of the kinetic energy spectrum. The lifetime of the localized electrons was ca. 1.6 ps (eventually these electrons recombine with the mirror charge). Initially, the electron is in an extended state with effective mass of 1.2 $m_e$. The localization proceeds with rate constants between $4 \times 10^{11}$ s$^{-1}$ and $3 \times 10^{12}$ s$^{-1}$ depending on the electron wave vector (band energy) that determines the exothermicity of localization (which varied between 10 meV and 150 meV). Both the thermalization dynamics and energetics were studied simultaneously. Hopefully, more ultrafast photoemission studies will follow. It would be particularly interesting to carry out such observations on small-diameter jets in order to study chemical processes in the bulk of the solvent. Studies on photoelectric spectroscopy of liquid jets (water, alcohols, *n*-nonane) were recently reported [75].

*3.2. Triplet state vs. singlet state formation: spin effects.* It is commonly believed that the initial spin state of the multiple-pair spur is singlet. Shortly after the ionization event, positively and negatively charged ions recombine yielding singlet or triplet products. In the contact exchange approximation, the multiplicity of the product depends on the spin phasing at the instance of recombination [32, 76]. (Calculations with more realistic spin exchange potentials support this approximation [77]) Recombination of the initially singlet-correlated ("geminate") ion pairs yields singlet products, including the fluorescent excited states. Random cross-recombination of ions in multiple-ion pair reduces the singlet yield and increases the triplet yield: statistically, 75% of the random-pair encounters yield the triplet excited states [76,78]. In squalane, the probability $f_S$ of the singlet recombinations decreases from 80% at 16 eV photoexcitation to less than 40% at 120 eV photoexcitation [32]. For alkanes, the most probable energy loss is 22-24 eV [79] and the steep decrease in $f_S$ with the photon energy follows the increasing formation of multiple-pair spurs.

Coherent spin evolution in the geminate ion pairs is the basis for several optical spectroscopies, including magnetic field effect [32,33,62-65,71,76,80], ODMR [24,39-48], magnetic level-crossing [21,28,50,59] and quantum beat spectroscopies [29-31,36,80,81]. Unpaired electrons in radical ions interact with static and microwave magnetic fields of the spectrometer (Zeeman interaction) and local magnetic fields generated by protons (hyperfine interactions). These weak magnetic interactions flip the electron spin and cause intersystem crossing in geminate pairs, eventually changing the singlet/triplet yields [76]. These spin-sorting interactions underline the magnetic and microwave field effects in radiolysis of hydrocarbons.

The importance of the magnetic and spin effects in radiolysis has been frequently discussed and we refer the reader to reviews by Brocklehurst [32,76]. There is one essential point which is seldom emphasized in the literature: Due to the very nature of quantum entanglement, the triplet and singlet channels exist *only in the context of the product formation.* In the absence of spin-sorting reactions or special initial conditions, *pairwise* spin correlations (i.e., "geminate" pairs) in many-particle spurs are forbidden by the laws of quantum mechanics. The only exceptions are single-pair spurs and spurs in which all but one pair have recombined. Given that the ion escape yield in most of hydrocarbons is < 5 %, the latter type includes most of the multiple-pair spurs that remain a few nanoseconds after the ionization event. Put together, these two types of spurs contain a large fraction of the long-lived "geminate" pairs formed in radiolysis and account for most of spin correlations observed in magnetic field and magnetic resonance experiments. As was noted by Brocklehurst [32], these long-lived pairs have a tendency to be singlet-correlated since most of the stable neutral products are singlet.

The origin of the spin effects in radiolysis has always been murky. The photoionization of the solvent could proceed directly or via short-lived highly excited solvent states. This autoionization may completely change the spur chemistry. A recent study examined the magnetic field effect on the solute fluorescence in squalane as a function of the VUV photon energy [32,62]. The magnetic field effect increased between 11 eV and 16 eV and reached the maximum corresponding to 80% singlet recombination. This probability is significantly lower than the value of 100% expected for single-pair spurs, while the photon energies do not allow one to account for the loss of spin coherence through cross-recombination. Analogous results were obtained for other systems (alcohols, benzene) [32]. Since ODMR and related studies indicate that for squalane holes the spin-lattice relaxation is longer than their decay [24,28,29], this loss of spin coherence cannot be accounted for by spin randomization of primary pairs due to magnetic interactions in the charges. It appears that the *initial state of the electron-hole pair is not purely singlet.* According to Brocklehurst [62], only direct ionization yields singlet-correlated electron-hole pairs while the autoionization results in the partial loss of spin coherence due to spin-orbital coupling in the highly excited states. This scenario is not entirely unrealistic given the possible excitonic nature of these states. However, it must be noted that the spin mixing postulated by Brocklehurst was not supported by VUV studies on solvent fluorescence [82].

Another fundamental problem is the wavefunction structure of the initial singlet state. Consider a spur consisting of two pairs, $(e_1h_1)$ and $(e_2h_2)$. There are two orthogonal singlet states, $|1\rangle=|e_1h_1\rangle_S |e_2h_2\rangle_S$ and $|2\rangle=1/\sqrt{3} \{|e_1h_2\rangle_S |e_2h_1\rangle_S + |e_1e_2\rangle_S |h_1h_2\rangle_S\}$, where $|..\rangle_S$ is the singlet state of the pair [77]. In the general case, the initial singlet state is a linear combination of $|1\rangle$ and $|2\rangle$. It is easy to demonstrate that pairwise correlations are possible only if the initial singlet function is multiplicative (e.g., state $|1\rangle$) [77]. In order to have these *pairwise* correlations *prior* to the spin-sorting recombination, the electron-hole pairs must be generated in spatially- and electronically-separated events [76,78]. While this assumption seems plausible in radiolysis of gases, in the condensed phase the excitation and charge delocalization could lead to significant mixing of the singlet wavefunctions. Another concern is that ionization events induced by low-energy secondary electrons entangle these electrons with the electron-hole pairs. Such entanglements create complex spin correlation patterns in the spur.

If the initial multiplicity of the spur is singlet, the probability $f_S$ to form the singlet product is given by the formula [78]

$$f_S \approx \Theta + 1/4 \, (1-\Theta), \qquad (16)$$

where $\Theta$ is the probability of recombination of "spin-correlated geminate pairs". This expression has been used to calculate the $f_S$ from the data on magnetic field effects [32] and quantum beats [31,80,81] in delayed solute luminescence (spin-coherence spectroscopies). Given that the premises of these methods are identical, it is understandable that the probabilities $f_S$ determined for fast electron spurs using these spectroscopies are comparable.

A serious complication in the measurements of $\Theta$ is the occurrence of spin relaxation in the charge carriers [33,50]. While the electrons have relatively long spin-lattice relaxation times, the solvent holes may relax on nanosecond or even subnanosecond time scales (see section 2.2). For ion pairs involving such species, the spin coherence is lost on the time scale of the singlet product formation. This problem is exacerbated when the geminate recombination is slowed down after scavenging of the primary ions by the scintillator. Even relatively slow spin relaxation induced by electron dipole-dipole coupling and anisotropic magnetic interactions in the radical cations is capable of destroying the initial spin correlation on the submicrosecond time scale [33]. This effect must be taken into account in the studies on long-lived ion pairs in viscous solvents and solids.

The common problem in studies on the efficiency of triplet and singlet recombinations is that only a fraction of the singlet/triplet products are detected. Alkane solvent holes rapidly fragment or transfer a proton which causes the loss of the solvent/solute luminescence. The recombination of pre-thermalized holes could yield other products than the lowest excited states [37,53]. In dense spurs, excited states can be quenched by radicals and radical ions.

20.

Computations show that these quenching reactions can be a significant mechanism of disappearance of the singlet excited states [65], and experimental evidence has been obtained supporting the reduced production of *solute* $S_1$ states in spurs of 20-100 eV photons due to the quenching [32,62,65]. Shortening of the *solvent* $S_1$ state lifetime with increasing excitation photon energy was observed in radiolysis of *cis*-decalin and *n*-dodecane with 4-to-14 keV x-rays [83]. This shortening was accounted for by quenching of the alkane $S_1$ states by alkyl radicals generated in the same spurs.

Even more important are reactions of radiolytic products with the *precursors* of these excited states. Results from recent experiments in which the solute luminescence and magnetic field effect were compared for radiolysis of cyclohexane or isooctane with 0.5-2.2 MeV electrons, 1-5 MeV protons, and 2-20 MeV $\alpha$-particles suggested that the decrease in the solute luminescence and the magnetic field effect was due to both the increasing importance of cross-recombination and the "intervention of radicals or other transient species with the precursors" with the fluorescent states [63]. The effects of spin relaxation and ion-radical reactions in dense spurs were identified as likely causes for reduced magnetic field effects, fluorescence yields, and probabilities $\Theta$ in spurs from 17-40 keV x-rays as compared to the spurs from fast electrons [80].

In short, to model the *experimentally-determined* singlet-to-triplet ratio, the initial state multiplicity, spur chemistry, and spin relaxation must be given same attention as the interplay between the cross- and geminate recombination.

*3.3. Modeling of light -particle spurs.* Many simulations of the spur kinetics in hydrocarbons have been reported over the last decade [84-89] (see reference [84] for a review). As may be anticipated, none of these simulations attained the level of complexity needed to obtain a self-consistent picture of the early radiolytic events in hydrocarbons. The threefold problem is the uncertainty about the initial spur structure, intractability of the inhomogeneous kinetics and poor understanding of spur chemistry. Due to the limited state of knowledge, the present-day simulations should be viewed as exploratory. While some researchers focus on the general aspects of spur dynamics using stochastic Monte Carlo simulations [84-88], others focus on the chemical kinetic aspects assuming simple (inhomogeneous) dynamics [23,24,89]. There are also models intended to integrate the chemical kinetics and the charge recombination dynamics [25,65]. At the present time, such models are ahead of their time, since only a few of the postulated 25-120 spur reactions have been studied experimentally.

Not having enough space to summarize all of the recent work, we will give a couple of examples. A welcome new development is the integration of energy-loss calculations [79,87]

21.

with the stochastic Monte Carlo simulations of charge recombination in order to obtain free ion yields as a function of the primary electron energy [87]. In these calculations, the charge carriers are treated as point charges of given mobility that migrate in their mutual Coulombic field. These simplifications are justifiable since the calculations of the track structure (which yield the initial positions of the point charges) are even more simplistic. The estimated free ion yields and their energy dependence [84,85,87] correspond well to those determined experimentally for spurs from soft x-rays [90]. The import of these calculations is that the distance distribution for thermalized electron-hole pairs is independent of the primary electron energy in the keV range, while ca. 50% wider distributions are needed to account for the free ion yields for MeV electrons [87]. This unexpected result needs further verification. There is also a report on the calculations of the distance distribution in the modified model of Mozumder and Magee [88]. It was found that in single-pair spurs the distribution was much more diffuse than in the multiple-pair spurs. The significance of this result for the free ion yield calculations has not been addressed.

A detailed chemical model of hydrocarbon radiolysis was advanced in order to account for the radiolytic yields of H atoms and $H_2$ as a function of iodine concentration (the latter was used as a radical scavenger) [89]. To simplify the kinetic scheme, it was assumed that the electron-hole pairs recombine yielding only the lowest singlet and triplet states of the solvent. Although this assumption is thoroughly unrealistic (and large corrections were made to recover the balance of radiolytic hydrogen), it allows simplification of the reaction scheme to just 10 reactions, using the deterministic diffusion-kinetic approach. Time-resolved yields of H atoms, $H_2$, radicals, olefins, and other products were estimated. From the simulation of the experimental data it appears that while the solvent $S_1$ states decay mainly via reaction (6) (the intersystem crossing was estimated to be only 10-15% efficient), the $T_1$ states decay via reactions

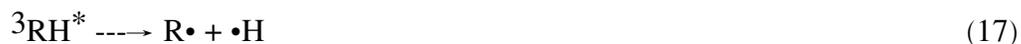
$$^3RH^* \longrightarrow R\bullet + \bullet H \tag{17}$$

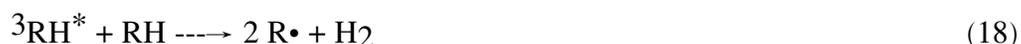
$$^3RH^* + RH \longrightarrow 2\,R\bullet + H_2 \tag{18}$$

with relative yields between 1:1 and 3:1, depending on the cycloalkane structure.

While being useful exercises, both of these simulations were based on simplified kinetic schemes and physical models of spur dynamics. For some problems (free ion yields) these considerations may be of little importance, for other (product yields) the oversimplification leads to ambiguous results. In particular, this applies to calculations of the "singlet yields" (implicitly identified with the yields of the lowest $S_1$ states) where the estimates entirely depend on the kinetic scheme [85,86]. Most of such calculations address only one facet of the problem: the

22.

relative significance of "geminate" recombination and cross-recombination in multiple-pair spurs. In the recent Monte-Carlo simulations [85], the initial spatial distribution of ionization events was calculated using the same approach that was used in the calculations of the free ion yield (see above). The total yield of singlet recombination for the primary pairs was estimated.

Superficially, these estimates correlate well with the energy dependence of the singlet ratio $f_S$ determined from the magnetic field effects on the solute luminescence yield [32,64]. However, some of the latter results were obtained using low-energy photons (< 120 eV) for which the stochastic Monte-Carlo calculations completely failed to reproduce the photon energy dependence of the free ion yield [64]. Radiolytic yields of the solvent $S_1$ states and free ions upon excitation of decalins and *n*-dodecane using 4 keV to 15 keV photons have been measured [83,90]. Even in this energy regime (where the Monte-Carlo calculations gave good estimates of the free ion yield), agreement between the calculated and measured yields was observed for one solvent only, *cis*-decalin. In the same studies, significant variation of the $S_1$ state lifetimes with the x-ray energy was observed which was a signature of quenching reactions in multiple-pair spurs [83]. Thus, the Monte-Carlo model was inadequate and the good correlation obtained between the experimental and simulated results was coincidental. Only when an understanding of the dynamics of short-lived excited states and pre-thermalized charges is reached, will development of an adequate model of the spur chemistry be possible.

## 4. GENERATION OF SOLVENT AND SOLUTE EXCITED STATES.

*4.1. The lower excited states.* Though excited solvent states have been discussed throughout this chapter, it is worth repeating a few points. Only the lower excited states have been considered, the fluorescing singlet $S_1$ state [91] and the putative, dissociative, triplet $T_1$ state [92]. These excited states are produced mainly via charge recombination; the singlet states may also be produced by direct excitation of the solvent by secondary electrons and Cerenkov light [93]. The $S_1$ state lives between 0.5 ns and 2 ns; this lifetime is longer (3-5 ns) for long-chain paraffins [92,94]. In multiple-pair spurs, the $S_1$ state lifetimes are shortened due to quenching of the excited state by radiolytic products [83]. Eventually, the $S_1$ state either fragments via reaction (6) or crosses to another excited state, possibly $T_1$ [92] or $S_2$ [95]; the radiative decay is very inefficient (< 1%). For homologous hydrocarbons, the quantum yield of fluorescence depends linearly on the excited state lifetime. For $C_7$-$C_{17}$ paraffins, the radiative rate constant is ca. $1.3 \times 10^6$ s$^{-1}$, for alkylcyclohexanes - $7 \times 10^6$ s$^{-1}$ [94]. The fragmentation is temperature-dependent (with activation energy of 0.145-0.21 eV), while the crossing is temperature-independent [92, 95, 96]. The yield of the $S_1 \rightarrow T_1$ (or $S_1 \rightarrow S_2$) crossing for cyclo- and *n*-alkanes varies between 10% and 40%; for some branched alkanes (such as isopropylcyclohexane and 2,3-dimethylbutane) this yield approaches 50-60% [92].

Cyclohexane fluoresces in liquid, solid and vapor, with lifetimes 0.8 ns (298 K), 1.85 ns (at 225 K), and 2.07 ns (298 K), respectively (7.3 eV excitation) [95]. For liquid cyclohexane at 298 K, the rate constant of crossing is $3.6 \times 10^8$ s$^{-1}$, the activation energy and pre-exponential factor for the fragmentation are 0.135 eV and $1.6 \times 10^{11}$ s$^{-1}$, respectively, and the quantum yield of fluorescence is ca. 0.01 at 7 eV photoexcitation [92]. For cyclohexane (and most of other saturated hydrocarbons), this quantum yield rapidly decreases with the excitation energy (to $3.5 \times 10^{-3}$ at 8.4 eV, the onset of photoionization) [60,82,92,94]. All of the hydrocarbon $S_1$ states have energies around 7 eV above the ground state (bottom-to-bottom) [92] and radiate between 5 eV and 6.5 eV [91]. The excited singlet states mainly dissociate to olefin/carbene + $H_2$ or alkyl radical + H atom. For cyclohexane (*n*-hexane), the relative yields of these two fragmentation channels change from 4:1 (3:1) upon 7.6 eV photoexcitation to 2:3 (1:1) upon 11.6 eV photoexcitation [97].

The nature of the long-lived $S_1$ states has been the subject of much speculation. Some authors view these states as liquid-phase analogs of Rydberg states in the gas phase. Others identify these states with self-trapped excitons, drawing from solid-state physics. Both of these descriptions offer no insight in what defines the experimental properties of these states, such as their reactions, energetics, and absorption spectra. Apparently, only first-principle calculations and further spectroscopic studies (in particular, in the IR) will be useful.

The lowest triplet states lay ca. 1 eV below the $S_1$ states [92]. These states are presumed to be extremely short lived (0.1-1 ps). The putative triplets are believed to decay via reactions (17) and (18). In paraffins and branched alkanes, the C-C bond dissociation is thought to be as efficient as the C-H bond dissociation; in alkylcyclohexanes, biradical formation was suggested [92]. Another study suggested the involvement of the $S_2$ state that dissociates yieding the same products as reaction (17) [95].

While this much is presumed or accepted, the role of these (and other) excited states in the radiolysis remains poorly understood. The results obtained over the last decade only add to the existing uncertainties.

*4.2. Non-fluorescing excited states of alkanes.* Several relatively long-lived non-fluorescing ("dark") excited states of hydrocarbons capable of sensitizing singlet solute luminescence have been proposed in recent years.

The strongest evidence of such states is from the work of Lipsky and co-workers [98-100] who have shown that the sensitization of solute fluorescence by transfer of energy from the photoexcited solvent (cyclohexane, *n*-hexadecane, 2,3-dimethylbutane, and *cis*- and *trans*-decalins) can be explained by the participation of *two* solvent excited states: the fluorescent $S_1$ state and a "dark" state. In *cis*-decalin containing 0.01 M of 2,5-diphenyloxazole (PPO), ca 10% of the solute luminescence is sensitized from this second singlet state. In a 20 µM solution, ca. 70% of the solute excitations are mediated by the "dark" state, with 0.25% of such energy transfers per each 7.7 eV photon absorbed or 0.06 per "dark" state produced. In 0.01 M solutions, the "dark" state transfers energy with 97% efficiency. In experiments where a biphotonic 3.7 eV or 4 eV laser excitation was used instead of 7.6 eV light, the same "dark" states were observed [99]. These states have a quantum yield of only a few percent of the fluorescent states and do not yield the $S_1$ states on deactivation. From the laser experiments, the conclusion was made that the "dark" states live much longer than the $S_1$ states, though their lifetimes must be less than 10 ns. Neither this lifetime nor the efficiency of the energy transfer from the "dark" state varied with the solvent viscosity when *cis*-decalin was diluted with higher-IP alkanes (2,3-dimethylbutane and isooctane). Lipsky and co-workers suggested that the "dark" states are exciplexes that migrate with a diffusion constant of $4 \times 10^{-4}$ cm$^2$/s [98]. Because the "dark" states have relatively low yield, their contribution to the solute fluorescence is most noticeable at low solute concentration.

A "dark" state capable of rapid energy transfer to aromatic solutes has been proposed in irradiated cyclohexane on the basis of experiments in which the formation of the $S_1$ state of *cis*-decalin (added as a scintillator) was measured and compared with the amount of this $S_1$ state expected from ion recombination and energy transfer from the $S_1$ state of cyclohexane [101]. A significant yield (0.8 per 100 eV) of the *cis*-decalin $S_1$ state was attributed to energy transfer from the "dark" state of cyclohexane. If one considers concentrated solutions, this "dark" state seems to be more important in the radiolysis than the "dark" state identified by Lipsky and co-workers. "Dark" exciton states were also postulated to account for the high yield of luminescence from 2,5-bis-(*tert*-butylbenzoxazolyl-2)-thiophene in picosecond pulse radiolysis of decalins and cyclohexane [102]. The radiolytic yield of this "dark" excitonic state was estimated at 1 per 100 eV and the rate of the energy transfer from these states was estimated as $(1-3) \times 10^{11}$ M$^{-1}$ s$^{-1}$. We are inclined to think that the observations of "dark" states in radiolysis are artifacts of data analysis. It remains to be seen whether these "dark" states are real.

*4.3. Low yields of the solvent $S_1$ states.* Radiolytic yields of the solvent $S_1$ states observed in the fast electron radiolysis of some alkanes (e.g., $C_3$-$C_8$ paraffins) and cycloalkanes (e.g., $C_5$-$C_7$ cycloalkanes and methylcyclohexane) are unexpectedly low (see tables IX.3-11 in reference [102] for available data). For example, in radiolysis of cyclohexane, the yield of the $S_1$ states is only 1.45±0.15 per 100 eV [104] (other estimates are 1.5±0.4 [105], 1.4-1.7 [100], and 1.75 per 100 eV [89]) while the $T_1$ states are generated at 3.4 per 100 eV [89], as estimated from the solvent luminescence yield and the product analysis, respectively. If these $S_1$ states were mainly produced in recombination of electron-hole pairs (as follows from studies of the effect of electron scavengers on the yield of solvent luminescence), the radiolytic yields would seemingly be equal to the G-value of ionization times the singlet recombination probability, $f_S$. For $f_S \approx 0.5$ (see below), one obtains the ionization yield of 3 pairs per 100 eV, whereas the experimental estimates are between 4.5 and 5 pairs per 100 eV [1]. Therefore, it appears that only 50-65% of the ionizations produce the electron-hole pairs that involve solvent radical cations.

One way to explain this deficit of solvent $S_1$ states is to postulate the prompt formation of "satellite ions" whose recombination does not yield the solvent $S_1$ states or any other fluorescent states [37,53]. As discussed in section 2.4, there are many results suggesting significant yield of "satellite ions" in ionization of hydrocarbons. For cyclohexane, the dc conductivity data suggest that ca. 50% of the cations observed 10 ns after the ionization event are not solvent holes. Promptly-generated cyclohexene$^{•+}$ ions were observed by ODMR and transient absorbance spectroscopies. In the latter experiments, the "satellite ions" were observed as early as 250 ps after the ionization event. Very high yields of fragment cations in cyclohexane cannot be accounted for by scavenging reactions (7) and (8) in radiolytic spurs. For *cis*- and *trans*-decalins, where no conductivity or magnetic resonance data exists that indicate a high yield of the "satellite ions", the radiolytic yield of the $S_1$ states is roughly twice that in cyclohexane (see below). Apparently, in some hydrocarbons there must be rapid fragmentation of pre-thermalized solvent holes. Cyclohexane is prone to such fragmentations: In gamma radiolysis of gaseous cyclohexane, the radiolytic yields of $C_6H_{12}^{•+}$, $C_6H_{11}^+$ (-H), and $C_6H_{10}^{•+}$ (-H$_2$) cations are 2.0, 0.62, and 0.35 species per 100 eV, respectively, out of the total ion pair yield of 4.4 per 100 eV [1]. In scavenging experiments with ammonia, the radiolytic yield of *c*-$C_6H_{11}^+$ cations in liquid cyclohexane was estimated as 0.7 ions per 100 eV [106].

Measurement of the effect of electron scavengers on the $^1RH^*$ yield show that scavengers are more effective in reducing this yield than they are in scavenging electrons. A kinetic mechanism was proposed in which a fraction of the solvent holes are initially in an excited state that does not yield the solvent $S_1$ state upon recombination [53]. The relaxation and fragmentation of the pre-thermalized solvent hole is expected to occur on the time scale of a few picoseconds.

26.

*4.4. Estimates of excited state generation by final product analysis.* We have already mentioned the product analysis studies on the yield of the solvent excited states and the products of their decomposition [89]. Using iodine scavenging, the radiolytic yields of $H_2$, olefins and radicals were determined for $C_5$-$C_8$ cycloalkanes as a function of LET [89,107]. To obtain the observed yield of $H_2$, a high yield of promptly generated $H_2$ (0.75 molecules per 100 eV for cyclohexane) was postulated. As discussed above, this hydrogen is likely to be from reaction (6). Therefore, it is instructive to compare the data on cyclohexane with such data for decalins, which seem to yield more stable excited solvent holes. Decalins also yield relatively stable highly excited states, since quantum yields of fluorescence vary little with the photon excitation energy [82]. This suggests efficient relaxation to the lowest $S_1$ state.

A recent study has provided the needed information [108]. Product yields from the $S_1$ state of *cis-* and *trans-* decalins were determined in 7.6 eV photolysis. Only 40-50% of these states were found to fragment (which is an unusually low dissociation yield for a saturated hydrocarbon). The yields of the solvent $S_1$ states in 3 MeV β-radiolysis were determined and, combined with the photolysis results, were used to estimate the fraction of radiolytic products that originate from the $S_1$ states. The rest of the products were assigned to the $T_1$ states. Table 1 summarizes the results. As seen from this table, for *cis-* and *trans-*decalins the probabilities $f_S$ of singlet recombination are close to 0.5. This probability is in fair agreement with the recently corrected estimate of $f_S \approx 0.65$ obtained from the measurements of absolute fluorescence yields [109]. According to Table 1, the $S_1$ yields in decalins are roughly twice that in cyclohexane while the $T_1$ yields for all three of the cycloalkanes are comparable. We conclude that highly excited states of neutral and charged decalins, unlike those of cyclohexane and many other hydrocarbons, exhibit low yields of fragmentation.

**TABLE 1.**

Radiolytic yields of solvent excited states in decalins (3 MeV electrons) [108] and cyclohexane ($^{60}Co$ γ-radiolysis) [89] estimated from product analysis.

| yield per 100eV | *cis-*decalin | *trans-*decalin | cyclo-hexane |
|---|---|---|---|
| $S_1$ state | 3.4 | 2.8 | 1.75 |
| $T_1$ state [a] | 3.0 | 3.2 | 3.40 |
| $f_S$ [b] | 0.53 | 0.47 | 0.34 |
| prompt $H_2$ | c | c | 0.75 |
| H$_2$ from S$_1$ | 1.43 | 1.40 | 1.49 |
| H$_2$ from T$_1$ | 2.47 | 3.00 | 1.9 d |

a) before S$_1$→T$_1$ conversion [89];

b) probability of singlet recombination for electron-hole pairs;

c) not included in the kinetic scheme;

d) reaction (18) only.

While product analysis is an important source of information on the early stages of radiolysis, this approach has a serious problem: For all of the alkanes studied, the yield of primary decomposition estimated from the total product yields significantly exceed the ionization yield: 6 to 6.5 per 100 eV vs. 4.5 to 5 per 100 eV [1]. This suggests that some of the products were counted twice: i.e., that fragmentation of the excited states is more extensive than was assumed. For example, instead of H$_2$ elimination, the exited state may eliminate two H atoms. Such decompositions were observed in gaseous methane (see discussion in reference [110]). In the fast-electron radiolysis of dilute solutions of methane in liquefied noble gases, the prompt yield of free H atoms is ca. 5 times higher than the prompt yield of methyl radicals. Using time-resolved EPR, these species were detected several tens of nanoseconds after their generation, and the yield of the methyl radicals was unaffected by cross-recombination (extremely long thermalization distances in the simple liquids make the kinetics homogeneous from the onset [110]). These results suggest that in a large fraction of recombination/excitation events involving methane and its cations, a methylene and two H atoms are generated. The implications of such fragmentations in radiolysis of neat hydrocarbons are obvious.

Alkyl radicals are one of the most chemically-important products of hydrocarbon radiolysis [1-3,60]. Their generation is very fast; e.g., for cyclohexyl radicals, the rise time of the formation is < 20 ps [22]. It is generally believed, that the radicals are generated in dissociation of the T$_1$ states or higher singlet states, in rapid reactions of "hot" H atoms and in "slow" reactions of thermalized H atoms generated in reaction (16) with hydrocarbons (the relaxed H atoms have lifetime of ca. 10 ns). Other pathways include deprotonation of solvent holes (reactions (1) and (4)) and neutralization of proton adducts and carbonium ions [37]. Radiolytic yields of radicals were estimated from the product analysis: the yield of cross-linking products, halocarbons generated upon scavenging the radicals with TI and I$_2$, and from the isotope sampling [1-3]. For long-chain paraffins, this analysis is complicated by uncertainties about the significance of hydrogen β-shifts and radical disproportionation [111]. Promptly generated radicals (< 100 ns) may be observed using time-resolved EPR; on this short timescale the secondary radical reactions are unimportant and one can directly measure the absolute and relative yields of neutral radicals [112,113]. Interestingly, these prompt yields change little upon addition of electron scavengers [113], which suggests that radicals might be generated on time scales faster than charge recombination (e.g., via reaction (4)).

For paraffins, the ratio of yields of penultimate and interior -H radicals is 20% higher than expected from the abundance of the corresponding C-H bonds (this was observed using both by the product analysis [111] and EPR [112]). Since the C-H bond dissociation energies for carbon-2 bonds are higher than those for interior carbon atoms, this preference cannot be explained through the abstraction by H atoms. One way to explain the observed preference is to assume the occurrence of reaction (4): it is known that in low-temperature solid paraffins reaction (1) yields mainly terminal and penultimate radicals [114]. In addition to the -H radicals, paraffins exhibit high yield of radicals formed upon C-C scissions, preferably for interior carbon atoms [3].

For branched alkanes, the fragmentation patterns could be very complex [3, 111]. The prompt yield of the -H radicals is always minor; the highest yields are of the radicals formed by scission of skeletal C-C bonds next to the branches. For example, in radiolysis of isooctane only 15% of the radicals are of the -H type, the rest being *tert*-butyl and 2-propyl radicals [111]. In radiolysis of 2,3-dimethylbutane, 70% of radicals are 2-propyl and 30% are -H radicals (2,3-dimethyl-2-butyl). It is not known what species dissociate (singlets? triplets? excited holes? excited radicals?) and what controls these fragmentation patterns.

*4.5. Generation of solute excited states.* In radiolysis of hydrocarbon solutions of aromatic scintillators (A), a significant fraction of solute fluorescence in the first several nanoseconds is sensitized by the energy transfer from the solvent $S_1$ state

$$^1RH^* + A \longrightarrow RH + {}^1A^* \tag{19}$$

Direct excitation of A is usually unimportant, but excitation of A by Cerenkov light is not negligible [49]. For many solutes, reaction (19) has large reaction radii (1-1.5 nm) and low activation energy (~ 50 meV).

Solute fluorescence is also induced by charge scavenging followed by radical ion pair recombination

$$e^- + A \longrightarrow A^{\bullet -} \tag{20}$$

$$RH^{\bullet +} + A \longrightarrow RH + A^{\bullet +} \tag{21}$$

$$RH^{\bullet +} + A^{\bullet -} \longrightarrow RH + {}^{1,3}A^* \tag{22}$$

29.

$$A^{\bullet+} + A^{\bullet-} \longrightarrow A + {}^{1,3}A^* \tag{23}$$

$$A^{\bullet+} + e^- \longrightarrow {}^{1,3}A^* \tag{24}$$

Triplet solute states ($^3A^*$) are formed only in these reactions. Since, typically, electron scavenging reaction (20) is faster than hole scavenging reaction (21), reactions (22) and (23) account for most of the delayed solute fluorescence. In the first few nanoseconds, only reactions (19) and (22) are important. Reaction (24) is important only in the studies on solute photoionization in hydrocarbons. The $^3A^* + {}^3A^*$ annihilation is a minor source of delayed fluorescence, though there have been reports to the contrary.

Radiolytic yields of solute $^1A^*$ states have been determined in several laboratories by measuring the fluorescence from solutions of aromatic scintillators on the subnanosecond time scale [49,100,101,109,115-118]. At Argonne [49,118], the motivation was to compare the experimental values with stochastic Monte-Carlo simulations using a single ion-pair model in order to determine whether the experimental kinetics could be matched by this model. The approximate correspondence between the experimental and simulated kinetics appeared to be possible only by making the assumption that the solvent holes were unstable on the time scale of the fluorescence measurement, undergoing a transformation that partially "disabled" reaction (22). This assumption allowed fair agreement with the *shapes* of the fluorescence kinetics for 1, 10 and 50 mM solutions of the aromatic scintillator, but the agreement with the absolute radiolytic yields was poor, and the magnitude of the disagreement varied with solute concentration.

A factor which contributes to the complexity of the analysis is the occurrence of reaction (19) which accounts for a large fraction of the $^1A^*$ states. For cyclohexane, this process can be accounted for with reasonable certainty. For *n*-hexane, the simulation is less satisfactory because the yield of the solvent $S_1$ state has not been accurately measured (1.6±0.5 per 100 eV [119]), and the estimates of its lifetime vary from 0.3 ns [95] to 0.7 ns [94].

Here we examine again what is perhaps the most puzzling aspect of the results, i.e., rapid leveling of the $^1A^*$ yield in dilute scintillator solutions (~ 1 mM). Qualitatively, simulations show that production of $^1A^*$ by energy transfer from the solvent $S_1$ state has a time profile very similar to the observed kinetics. The question is why does not the yield of $^1A^*$ continuously increase after 2 ns despite the occurrence of reactions (21) and (22)? Since the cyclohexane hole has long lifetime, the flatness of the kinetics must be due to something else than the hole instability. For *n*-hexane, a re-examination of the results indicates that the occurrence of reaction (22) is needed to explain the kinetics observed; the latter can be simulated assuming that 50% of recombinations (22) yield $^1A^*$. However, the calculated G($^1A^*$) at 5 ns is larger than the experimental value of 0.07 per 100 eV by a factor of $\approx 1.8$, so quantitatively, the situation with *n*-hexane is not settled.

While it is an open question whether the quenching reactions in multiple-pair spurs and sensitization of solute fluorescence by "dark" states can explain the discrepancies, it seems more likely that the loss of the solute luminescence is due to some irregularity in the behavior of cyclohexane holes. One possibility is that rapid electron spin-relaxation in these holes randomizes {RH•+ A•-} geminate pairs and reduces the $^1A^*$ yield (see section 2.2). The occurrence of such randomization does not contradict the experimental estimates of $f_S \approx$ (0.5-0.6) for recombination of secondary pairs: these estimates were obtained in concentrated solutions (>0.1 M) of the scintillator in which the solvent hole was scavenged prior to the spin relaxation (see below). Alternatively, one may speculate that for cyclohexane, reaction (22) does not produce solute excited states or at least does so with reduced efficiency.

*4.6. Time-resolved measurements of the singlet recombination probability.* For saturated hydrocarbons irradiated with fast electrons and γ-rays, the estimates of singlet recombination probability $f_S$ obtained using spin coherence methods (such as magnetic field effect and quantum beat spectroscopy) are between 0.4 and 0.65 [31,32,80,81]. These probabilities were estimated from the fraction $\Theta$ of spin-correlated geminate pairs by implementation of formula (16). Importantly, this fraction was determined for *secondary* radical-ion pairs. To prevent the loss of spin coherence due to spin relaxation in the solvent holes, both the solvent hole and the electron were scavenged in less than 1 ns, using 0.12 M diphenylsulfide and 1 mM *para*-terphenyl, respectively [80,81]. Still, the estimates of $\Theta$ were compromised by involvement of pairs that included "satellite" ions, such as olefin radical ions [81]. In the latter species, the electron spin is strongly coupled to protons. This speeds up the spin dephasing of the geminate pair. Thus, the reported values of $\Theta$ are the low-limit estimates [31].

Given that the validity of formula (16) and the premises of the spin-coherence methods are not obvious, it was important to determine $f_S$ for secondary ion pairs in a direct way. The time dependence of the $^1A^*$ and $^3A^*$ yields in 0.1 M solutions of biphenyl and naphthalene in cyclohexane, *n*-hexane, and isooctane was measured [118]. In these solutions, reactions (19), (20), and (21) was over in a fraction of a nanosecond, and reaction (23) was the only source of $^1A^*$ and $^3A^*$ states between 1 ns and 70 ns. From the derivatives of the G-values of $^3A^*$ and $^1A^*$, the time-dependent probability $f_t = 1 - f_S$ of triplet recombination was obtained. This probability was 0.5±0.1 regardless of the delay time. Surprisingly, no experimentally significant decrease in $f_S$ during the first 10 ns after the ionization event (due to spin mixing in

31.

the secondary geminate pairs, e.g., by hyperfine interaction in the radical ions) was found. Thus, the spin-coherence methods seem to give reliable estimates of $f_S$.

In the same work, the probability $f_t$ for long-lived secondary pairs was estimated by time-resolved measurements of G($^3$A$^*$) and G(A$^{\bullet-}$) over the solute concentration range 1 mM to 0.1 M and for times out to 1 µs. Corrections were made to take into account dimerization of A$^{\bullet+}$, $^1$A$^* \to {}^3$A$^*$ crossing, triplet-triplet annihilation, reactions of $^1$A$^*$ and A$^{\bullet-}$ with cyclohexyl radicals, etc. [118]. No provisions were made to account for spin dephasing in secondary pairs, though this dephasing is not negligible on the long time scale. It was found that the kinetics simulated using $f_t \approx 0.5$-$0.7$ were in a reasonable agreement with the experimental ones, and the fit quality was not improved by allowing $f_t$ to vary with time.

An interesting result that still awaits theoretical explanation is a remarkable similarity in the formation kinetics for $^1$A$^*$ and $^3$A$^*$ states observed between 60 ps to 5 ns in concentrated solutions of aromatic scintillators in *n*-hexane (0.05-0.1 M) [49,118]. At these concentrations, the electrons are scavenged in < 10 ps and the formation of the solvent S$_1$ states is much reduced, so that reaction (19) is relatively unimportant. Thus, most of the $^1$A$^*$ states are formed in the same reaction as the $^3$A$^*$ states, via recombination of the solvent holes (reaction (22)) and "satellite ions" with A$^{\bullet-}$. The similarity of the formation kinetics for the $^{1,3}$A$^*$ states suggests that on the short time scale the probability of cross-recombinations and geminate recombinations are very similar. It was concluded that "the spurs are made of a collection of negative and positive ions instead of separate geminate pairs" [118]. As was emphasized above, this view (the loss of pairwise spin correlations in multiple-pair spurs) does not contradict the occurrence of spin coherence phenomena in radiolysis. Whether this result can be explained using the conventional Monte-Carlo models of the spur kinetics needs to be determined.

## 5. CONCLUDING REMARKS

Radiation chemistry of saturated hydrocarbons is far from being well understood, and is a province of numerous controversies and speculations. No closure is in sight. At the same time, no other medium with exception of liquid water has been studied as comprehensively as liquid alkanes and cycloalkanes. Below, we provide our list of the most important and challenging problems in the radiation chemistry of saturated hydrocarbons:

- *The chemistry of "hot" intermediates:* pre-thermalized charges, highly excited solvent states, energetic fragments (such as "hot" H atoms). What is the nature of these states ? What role do they play in radiolysis? How do they relax and react ? *What happens to the heat dissipated in radiolytic reactions?* What is the mechanism for vibrational deactivation of the products? Could the heat be dissipated through the formation of unusual conformers and what chemical consequence would that have?

- *Understanding the spur structure* as a function of the electron energy, especially below 1 keV. What is the relative significance of ionization and autoionization processes? What is the origin of the observed spin effects?

- *First-principle simulations of the excitation dynamics, charge localization, and charge transport in liquid hydrocarbons.* Are there excitons, Rydberg states, and exciplexes in liquid hydrocarbons? What are the mechanisms for localization of electrons and holes in non-polar liquids? What is the mechanism for rapid diffusion of holes and excited states? What determines the fragmentation pathways of triplet and singlet excited states?

- *The role of "silent" species*, in particular, hydrogen atoms, "dark" states, proton adducts, and carbonium ions. How do these species form and interact with each other, excited states, and primary and secondary ions? Are there chemically-significant short-lived intermediates, such as carbenes and biradicals or some other unrecognized species?

This list may be expanded to include most of the issues discussed in this chapter. A lot of these problems stem from insufficient knowledge of short-lived intermediates and the complexity of spur dynamics. The use of traditional methods of radiation chemistry, such as product analysis or transient absorption and fluorescence spectroscopies, has its limitations. The "real action" in radiolysis takes place within the first few tens of picoseconds, when the hot species form, relax, fragment, and react. The existing pulse radiolysis facilities provide the time resolution to 20-30 ps. A new generation of laser-coupled linacs will push this time resolution to 500 fs [120]. The linac at the Osaka University already provides pulses of 800 fs with 2 nC charge (3 ps time resolution). Tabletop accelerators that use intense laser light ($> 10^{18}$ W/cm$^2$) to generate ~ 1 nC of photoelectrons with energies up to 30 MeV have been demonstrated [121,123]; with this technique, 50-100 fs electron pulses will be available. Thus, the time resolution will soon be improved. Will this improvement bring the closure?

Several decades ago, picosecond pulse radiolysis was as eagerly anticipated as the femtosecond pulse radiolysis today. Knowing what followed thereafter suggests that, by itself, the improved time resolution would not solve many outstanding problems: Though quite a few of the intermediates in radiolysis of hydrocarbons (e.g., solvent holes and solvent $S_1$ states) live for nanoseconds, their nature remains elusive and their formation and decay mechanisms undetermined. The most daunting task in the development of ultrafast pulse radiolysis is not only in the generation of short electron pulses. Rather, it is the development of better detection techniques (see reference [120], p. 17). Transient absorption spectroscopy was proven inadequate on the time scales of picoseconds or nanoseconds; there is no reason to expect that it would fare better on the shorter time scales. Actually, the situation would be worse since there are fewer sufficiently fast reactions to sort out light-absorbing species by their reactivity. Separating the overlapping absorbance signals from excited states, charge carriers, and fragments would be impossible given that none of these species have distinctive absorption bands. Most of techniques that complemented transient absorption spectroscopy on longer time scales (magnetic resonance, conductivity, fluorescence spectroscopy, etc.) cannot be used on the picosecond and subpicosecond time scales.

We conclude that putting the emphasis on the time resolution without offering adequate spectroscopic base would ensure a new stalemate. Therefore, *the development of fast, highly selective and sensitive techniques for detection of short-lived intermediates in spurs is the most urgent experimental problem in radiolysis of liquids.* Ultrafast laser spectroscopy must be utilized to prepare the ground for the ultrafast pulse radiolysis with the development of better detection techniques.

## REFERENCES

1. A. Hummel, in The Chemistry of Alkanes and Cycloalkanes, S. Patai and Z. Rappoport (eds.), John Wiley, New York, 1992, p. 743.

2. A. J. Swallow, in Radiation Chemistry: Principles and Applications, Farhataziz and M. A. J. Rodgers (eds.), VCH Publishers, 1987, p. 351.

3. R. A. Holroyd in Fundamental Processes in Radiation Chemistry, P. Ausloos (ed.), Interscience, New York, p. 413.

4. G. Beck and J. K. Thomas, J. Phys. Chem. 76 (1972) 3856.

5. A. Hummel and L. H. Luthjens, J. Chem. Phys. 59 (1973) 654.

6. E. Zador, J. M. Warman and A. Hummel, J. Chem. Phys. 62 (1975) 3897; Chem. Phys. Lett. 23 (1973) 363; J. Chem. Soc. Farad. Trans. I, 75 (1979) 914.

7. M. P. de Haas, J. M. Warman, P. P. Infelta and A. Hummel, Chem. Phys. Lett. 31 (1975) 382; ibid. 43 (1976) 321; Can. J. Chem. 55 (1977) 2249.

8. J. M. Warman, The Study of Fast Processes and Transient Species by Electron-Pulse Radiolysis; J. H. Baxendale and F. Busi (eds.), Reidel, The Netherlands, 1982, p. 433.

9.  J. M. Warman, H. C. de Leng, M. P. de Haas and O. A. Anisimov, Radiat. Phys. Chem. 36 (1990) 185.

10. I. A. Shkrob, A. D. Liu, M. C. Sauer, Jr., K. H. Schmidt and A. D. Trifunac, J. Phys. Chem. 102 (1998) 3363.

11. I. A. Shkrob, A. D. Liu, M. C. Sauer, Jr., K. H. Schmidt and A. D. Trifunac, J. Phys. Chem. 102 (1998) 3371.

12. A. D. Liu, I. A. Shkrob, M. C. Sauer, Jr. and A. D. Trifunac, J. Phys. Chem. 51 (1998) 273.

13. I. A. Shkrob, M. C. Sauer, Jr., K. H. Schmidt, A. D. Liu, J. Yan and A. D. Trifunac, J. Phys. Chem. 101 (1997) 2120.

14. M. C. Sauer, Jr., I. A. Shkrob, J. Yan, K. H. Schmidt and A. D. Trifunac, J. Phys. Chem. 100 (1996) 11325.

15. A. D. Liu, M. C. Sauer, Jr. and A. D. Trifunac, J. Phys. Chem. 97 (1993) 11265.

16. K. H. Schmidt, M. C. Sauer, Jr., Y. Lu and A. Liu, J. Phys. Chem. 94 (1990) 244.

17. M. C. Sauer, Jr. and K. H. Schmidt, Radiat. Phys. Chem. 32 (1988) 281.

18. M. C. Sauer, Jr., K. H. Schmidt and A. Liu, J. Phys. Chem. 91 (1987) 4836.

19. M. C. Sauer, Jr., A. D. Trifunac, D. B. McDonald and R. Cooper, J. Phys. Chem. 88 (1984) 4096.

20. R. Mehnert, in Radical Ionic Systems, Properties in Condensed Phase, A. Lund and M. Shiotani (eds.), Kluver, Dordercht, 1991, p. 231; R. Mehnert, O. Brede and W. Naumann, Ber. Bunsenges. Phys. Chem. 89 (1985) 1031; O. Brede, J. Bös, W. Naumann and R. Menhert, Radiochem. Radioanal. Lett. 35 (1978) 85.

21. F. B. Sviridenko, D. V. Stass and Yu. N. Molin, Chem. Phys. Lett. 297 (1998) 343.

22. S. Tagawa, N. Hayashi, Y. Yoshida, M. Washio and Y. Tabata, Radiat. Phys. Chem. 34 (1989) 503.

23. I. A. Shkrob, M. C. Sauer, Jr., J. Yan and A. D. Trifunac, J. Phys. Chem. 100 (1996) 6876.

24. I. A. Shkrob, M. C. Sauer, Jr. and A. D. Trifunac, J. Phys. Chem. 100 (1996) 5993; I. A. Shkrob and A. D. Trifunac, J. Phys. Chem. 100 (1996) 14681.

25. I. A. Shkrob, M. C. Sauer, Jr. and A. D. Trifunac, J. Phys. Chem. 100 (1996) 7237.

26. I. A. Shkrob, M. C. Sauer, Jr. and A. D. Trifunac, J. Phys. Chem. B 103 (1999) 4773; *ibid.*, B 104 (2000) 3752 and 3760.

27. V. I. Borovkov, S. V. Anishchik and O. A. Anisimov, in Proceedings of the 21st Miller Conference on Radiation Chemistry, April 24-29, 1999, Doorwerth, The Netherlands, p. 31.

28. B. M. Tadjikov, D. V. Stass, O. M. Usov and Yu. N. Molin, Chem. Phys. Lett. 273 (1997) 25.

29. O. M. Usov, D. V. Stass, B. M. Tadjikov and Yu. N. Molin, J. Phys. Chem. A 101 (1997) 7711.

30. A. V. Veselov, O. A. Anisimov and Yu. N. Molin, in Pulse Radiolysis, Y. Tabata (ed.), CRC Press, Boston, 1991, p. 27.

31. Yu. N. Molin, Bull. Korean Chem. Soc. 20 (1999) 7.

32. B. Brocklehurst, Radiat. Phys. Chem. 50 (1997) 213.

33. B. Brocklehurst, J. Chem. Soc. Farad. Trans. 93 (1997) 1079; M. Okazaki, Y. Tai, K. Nunome and K. Toriyama, Chem. Phys. 161 (1992) 177.

34. V. I. Borovkov, S. V. Anishchik and O. A. Anisimov, Chem. Phys. Lett. 270 (1997) 327.

35. A. Hummel, private communication.

36. V. M. Grigoryants, B. M. Tadjikov, O. M. Usov and Yu. N. Molin, Chem. Phys. Lett. 246 (1995) 392.

37. M. C. Sauer, Jr., D. W. Werst, C. D. Jonah and A. D. Trifunac, Radiat. Phys. Chem. 37 (1991) 461; A. D. Trifunac, M. C. Sauer, Jr. and C. D. Jonah, Chem. Phys. Lett. 113 (1985) 316.

38. A. D. Trifunac, M. C. Sauer, Jr., I. A. Shkrob and D. W. Werst, Acta Chem. Scand. 51 (1997) 158.

39. I. A. Shkrob and A. D. Trifunac, J. Phys. Chem. 98 (1994) 13262.

40. D. W. Werst, M. G. Bakker and A. D. Trifunac, J. Am. Chem. Soc. 112 (1990) 40.

41. A. D. Trifunac, D. W. Werst and L. T. Percy, Radiat. Phys. Chem. 34 (1989) 547.

42. D. W. Werst and A. D. Trifunac, J. Phys. Chem. 92 (1988) 1093.

43. D. W. Werst, L. T. Percy and A. D. Trifunac, Chem. Phys. Lett. 153 (1988) 45.

44. J. P. Smith, S. Lefkowitz and A. D. Trifunac, J. Phys. Chem. 86 (1982) 4347.

45. B. M. Tadjikov, N. N. Lukzen, O. A. Anisimov and Yu. N. Molin, Chem. Phys. Lett. 171 (1990) 413.

46. B. M. Tadjikov, V. I. Melekhov, O. A. Anisimov and Yu. N. Molin, Radiat. Phys. Chem. 34 (1989) 353.

47. V. I. Melekhov, O. A. Anisimov, V. A. Veselov and Yu. N. Molin, Chem. Phys. Lett. 127 (1986) 97.

48. V. I. Melekhov, O. A. Anisimov, V. A. Saik and Yu. N. Molin, Chem. Phys. Lett. 112 (1984) 106.

49. M. C. Sauer, Jr., C. D. Jonah and C. A. Naleway, J. Phys. Chem. 95 (1991) 730.

50. B. M. Tadjikov, D. V. Stass and Yu. N. Molin, J. Phys. Chem. A 101 (1997) 377.

51. A. Lund, M. Lindgren, S. Lunell and J. Maruani, in Molecules in Physcis, Chemistry, and Biology, vol. 3, J. Maruani (ed.), Kluwer, Dordrecht, 1989, p. 259; P. Wang, M. Shiotani and S Lunell, Chem. Phys. Lett. 292 (1998) 110, and references therein.

52. V. I. Melekhov, O. A. Anisimov, L. Sjöqvist and A. Lund, Chem. Phys. Lett. 174 (1990) 95.

53. C. D. Jonah and M. C. Sauer, Jr., Radiat. Phys. Chem. 46 (1989) 497.

54. D. B. Johnston, Y.-M. Wang and S. Lipsky, Radiat. Phys. Chem. 38 (1991) 583.

55. R. E. Bühler, Res. Chem. Intermed. 25 (1999) 259; R. E. Bühler and Y. Katsumura, J. Phys. Chem. A 102 (1998) 111.

56. Y. Katsumura, T. Azuma, M. A. Quadir, A. S. Domazou and R. E. Bühler, J. Phys. Chem. 99 (1995) 12814.

57. M. A. Lewis and C. D. Jonah, Radiat. Phys. Chem. 33 (1989) 1; B. C. Le Motais and C. D. Jonah, Radiat. Phys. Chem. 33 (1989) 505.

58. S. G. Lias, P. Ausloos and Z. Horvath, Int. J. Chem. Kinet. 8 (1976) 725.

59. D. V. Stass, N. N. Lukzen, B. M. Tadjikov, V. M. Grigoryants and Yu. N. Molin, Chem. Phys. Lett. 243 (1995) 533.

60. P. Ausloos, R. E. Rebbert, F. P. Schwartz and S. G. Lias, Radiat. Phys. Chem. 21 (1983) 27.

61. E. Haselbach and T. Bally, Pure Appl. Chem. 56 (1984) 1203 and references there.

62. B. Brocklehurst, Radiat. Phys. Chem. 50 (1997) 393.

63. J. A. LaVerne and B. Brocklehurst, Radiat. Phys. Chem. 47 (1996) 71; J. Phys. Chem. 100 (1996) 1682.

64. B. Brocklehurst, Chem. Phys. Lett. 211 (1993) 31.

65. B. Brocklehurst, J. Chem. Soc. Farad. Trans. 88 (1992) 167 and 2823.

66. H. Miyasaka and N. Mataga, in Pulse Radiolysis, Y. Tabata (ed.), CRC Press, Boston, 1991, p. 173; Radiat. Phys. Chem. 32 (1988) 177; Chem. Phys. Lett. 134 (1987) 480, ibid. 126 (1986) 219.

67. M. U. Sander, U. Brummund, K. Luther and J. Troe, J. Phys. Chem. 97 (1993) 8378.

68. F. H. Long, H. Lu and K. B. Eisenthal, J. Phys. Chem. 99 (1995) 7436.

69. L. D. A. Siebbeles, U. Emmerichs, A. Hummel and H. J. Bakker, J. Chem. Phys. 107 (1997) 9339.

70. G. Orlandi, L. Flamingi, F. Barigiletti and S. Dellonte, Radiat. Phys. Chem. 21 (1983) 113.

71. V. O. Saik and S. Lipsky, Chem. Phys. Lett. 264 (1997) 649; V. O. Saik, A. E. Ostafin and S. Lipsky, J. Chem. Phys. 103 (1995) 7347.

72. J. P. Dodelet, K. Shinsaka and G. Freeman, Can. J. Chem. 54 (1976) 744.

73. R. A. Crowell and D. M. Bartels, J. Phys. Chem. 100 (1996) 17940.

74. N.-H. Ge, C. M. Wong, R. L. Lingle, Jr., J. D. McNeill, K. J. Gaffney and C. B. Harris, Science 279 (1998) 202 [also see their review in J. Phys. Chem. B 103 (1999) 282].

75. M. Faubel, B. Steiner and J. P. Toennis, J. Chem. Phys. A 106 (1997) 9013.

76. B. Brocklehurst, Radiat. Phys. Chem. 21 (1983) 57; J. Chem. Soc. Farad. Trans. II, 72 (1976) 1869; Nature 221 (1969) 921.

77. C. E. Bolton and N. J. B. Green, J. Phys. Chem. 100 (1996) 8807.

78. W. M. Bartczak, M. Tachiya and A. Hummel, Radiat. Phys. Chem. 36 (1990) 195.

79. J. A. LaVerne and S. M. Pimblott, J. Phys. Chem. 99 (1995) 10540.

80. S. V. Anishchik, O. M. Usov, O. A. Anisimov and Yu. N. Molin, Radiat. Phys. Chem. 51 (1998) 31.

81. O. M. Usov, V. M. Grigoryants, B. M. Tadjikov and Yu. N. Molin, Radiat. Phys. Chem. 49 (1997) 237.

82. A. E. Ostafin and S. Lipsky, J. Chem. Phys. 98 (1993) 5408.

83. R. A. Holroyd, J. M. Preses and J. C. Hanson, J. Phys. Chem. A 101 (1997) 6931; Radiat. Res. 135 (1993) 312.

84. W. M. Bartczak and A. Hummel, J. Radioanal. Nucl. Chem. 232 (1998) 7.

85. L. D. A. Siebbeles, W. M. Bartczak, M. Terrisol and A. Hummel, in Microdosimetry. An Interdisciplinary Approach, D. T. Goodhead, P. O'Neill and H. G. Menzel (eds.), The Royal Society of Chemistry, Special Publ. No. 204, Cambridge, 1997, p. 11.

86. W. M. Bartczak and A. Hummel, Chem. Phys. Lett. 208 (1993) 232; Radiat. Phys. Chem. 39 (1992) 29.

87. L. D. A. Siebbeles, W. M. Bartczak, M. Terrisol and A. Hummel, J. Phys. Chem. A 101 (1997) 1619; W. M. Bartczak and A. Hummel, J. Phys. Chem. 97 (1993) 1253.

88. L. Musolf, W. M. Bartczak, M. Wojcik and A. Hummel, Radiat. Phys. Chem. 47 (1996) 83.

89. J. A. LaVerne, S. M. Pimblott and L. Wojnárovits, J. Phys. Chem. A 101 (1997) 1628.

90. R. A. Holroyd and T. K. Sham, Radiat. Phys. Chem. 51 (1998) 37; J. Phys. Chem. 89 (1985) 2909; R. A. Holroyd, T. K. Sham, B.-X. Yang and X.-H. Feng, J. Phys. Chem. 96 (1992) 7438.

91. F. Hirayama and S. Lipsky, J. Chem. Phys. 51 (1969) 3616; F. Hirayama, W. Rothman and S. Lipsky, Chem. Phys. Lett. 5 (1969) 296; W. Rothman, F. Hirayama and S. Lipsky, J. Chem. Phys. 58 (1973) 1300.

92. L. Wojnárovits and G. Földiák, Triplet States of Alkanes, in Proceedings of the 5th Working Meeting on Radiat. Interact., H. Mai, O. Brede and R. Mehnert (eds.), ZFI-Mitteilungen, Leipzig, 1991, p. 64.

93. M. C. Sauer, Jr., C. D. Jonah, B. C. Le Motais and A. C. Chernovitz, J. Phys. Chem. 92 (1988) 4099.

94. R. Hermann, R. Mehnert, and L. Wojnárovits, J. Lumin. 33 (1985) 69; Y. Katsumura, Y. Yoshida, S. Tagawa and Y. Tabata, Radiat. Phys. Chem. 21 (1983) 103.

95. M. A. Wickramaaratchi, J. M. Preses, R. A. Holroyd and R. E. Weston, Jr. , J. Chem. Phys. 82 (1985) 4745.

96. J. M. Preses and R. A. Holroyd, J. Chem. Phys. 92 (1990) 2938; S. Dellonte, L. Flamingi, F. Bargialetti, L. Wojnárovits and G. Orlandi, J. Phys. Chem. 88 (1984) 58.

97. F. P. Schwarz, D. Smith, S. G. Lias, and P. Ausloos, J. Chem. Phys. 75 (1981) 3300.

98. T. S. R. Krishna and S. Lipsky, J. Phys. Chem. A 102 (1998) 496.

99. Y.-M. Wang, D. B. Johnston and S. Lipsky, J. Phys. Chem. 97 (1993) 403 and references 3 to 10 therein.

100. L. Walter and S. Lipsky, Int. J. Radiat. Phys. Chem. 7 (1975) 175.

101. L. H. Luthjens, H. C. de Leng, L. Wojnárovits and A. Hummel, Radiat. Phys. Chem. 26 (1985) 509.

102. V. M. Grigoryants and V. V. Lozovoy, High Energy Chem. 30 (1996) 38.

103. Y. Tabata, Y. Ito, and S. Tagawa, CRC Handbook of Radiation Chemistry, CRC Press, Boston, 1991.

104. H. T. Choi, D. Askew and S. Lipsky, Radiat. Phys. Chem. 19 (1982) 373.

105. L. Wojnárovits and G. Földiák, Acta Chim. Acad. Sci. Hung. 105 (1980) 27.

106. T. Wada, S. Shida and Y. Hatano, J. Phys. Chem. 79 (1975) 561; S. Shida and Y. Hatano, Int. J. Radiat. Phys. Chem. 8 (1974) 171.

107. L. Wojnárovits and J. A. LaVerne, J. Radioanal. Nucl. Chem. 232 (1998) 19; J. Phys. Chem. 99 (1995) 3168; J. Phys. Chem. 98 (1994) 6046; J. A. LaVerne and L. Wojnárovits, Radiat. Phys. Chem. 47 (1996) 353; L. Wojnárovits and J. A. LaVerne, G. Földiák, M. Roder and L. Wojnárovits, Hung. Magy. Kem. Foly. 100 (1994) 184; J. Radioanal. Nucl. Chem. 165 (1992) 385; ibid. 164 (1992) 29.

108. A. Hummel, H. C. de Leng, L. H. Luthjens and L. Wojnárovits, J. Chem. Soc. Farad. Trans. 90 (1994) 2459.

109. L. H. Luthjens, H. C. de Leng, W. R. S. Appleton and A. Hummel, Radiat. Phys. Chem. 36 (1990) 213; L. H. Luthjens, P. Dorenbos, J. T. M. de Haas, A. Hummel and C. W. E. van Eijk, Radiat. Phys. Chem., in press.

110. I. A. Shkrob and A. D. Trifunac, Radiat. Phys. Chem. 50 (1997) 227; J. Phys. Chem. 99 (1995) 11122.

111. B. Tilquin and and T. Baudston, Radiat. Phys. Chem. 37 (1991) 23.

112. I. A. Shkrob and and A. D. Trifunac, Radiat. Phys. Chem. 46 (1995) 83.

113. D. W. Werst and A. D. Trifunac, Chem. Phys. Lett. 137 (1987) 475.

114. K. Toriyama, M. Iwasaki, Fukaya, H. Muto and K. Nunome, J. Chem. Soc. Chem. Commun. (1982) 1293; J. Phys. Chem. 89 (1985) 5278.

115. V. V. Lozovoy, V. M. Grigoryants, O. A. Anisimov, Generation of Excited States of Solutes in Radiolysis of Dilute Hydrocarbons, Inst. Chem. Kinet. & Combustion, Novosibirsk, Russia, Preprint 28, 1988 (in Russian); V. M. Grigoryants, V. V. Lozovoy, Yu. D. Chernousov, I. E. Shebolaev, V. A. Arutyunov, O. A. Anisimov and Yu. N. Molin, Radiat. Phys. Chem. 34 (1989) 349.

116. Y. Yoshida, S. Tagawa, M. Washio, H. Kobayashi and Y. Tabata, Radiat. Phys. Chem. 34 (1989) 493; ibid. 30 (1987) 83; Y. Katsumura, Y. Yoshida, S. Tagawa and Y. Tabata, Radiat. Phys. Chem. 21 (1983) 103; S. Tagawa, Y. Katsumura, and Y. Tabata, Radiat. Phys. Chem. 19 (1982) 125.

117. L. Wojnárovits, L. H. Luthjens, H. C. de Leng and A. Hummel, J. Radioanal. Nucl. Chem. 101 (1986) 349.

118. M. C. Sauer, Jr. and C. D. Jonah, Radiat. Phys. Chem. 44 (1994) 281.

119. L. Wojnárovits and G. Földiák, ZFI-Mitteilungen (Leipzig) 43a (1981) 243.

120. Research Needs and Opportunities in Radiation Chemistry, US-DOE Office of Science, BES-CHEM, Chesterton, Indiana, April 19-22, 1998, p. 13.

121. G. A. Mourou, C. P. J. Barty and M. D. Perry, Physics Today 51 (1998) 22; S. J. Matthews, Laser Focus World 35 (1999) 155.

122. I. A. Shkrob, A. D. Liu, M. C. Sauer, Jr., and A. D. Trifunac, J. Phys. Chem. B, in press.

123. N. Saleh, K. Flippo, K. Nemoto, D. Umstadter, R. A. Crowell, C. D. Jonah, A. D. Trifunac, Rev. Sci. Instr. 71 (2000) 2305.